\documentclass[longauth]{aa}
\usepackage[english]{babel}
\makeatletter
\usepackage[varg]{txfonts}
\usepackage{amsmath,mathtools,mathrsfs}
\usepackage{gensymb}
\usepackage{float}

\usepackage[colorlinks, allcolors=blue, breaklinks=true]{hyperref}
\addto\extrasenglish{%

}
\usepackage{hyperref}
\usepackage{verbatim}
\usepackage{color}
\usepackage{xspace}
\usepackage{longtable}
\usepackage{multirow}
\usepackage{afterpage}
\usepackage{booktabs}
\usepackage{bm}
\usepackage{subdepth}
\usepackage{amsmath}
\usepackage[caption=false]{subfig}
\usepackage[normalem]{ulem}
\usepackage{tabularx}
\usepackage{bm}
\usepackage{xcolor}
\usepackage{adjustbox}
\usepackage{array}
\usepackage[normalem]{ulem}
\usepackage[utf8]{inputenc}
\usepackage[switch]{lineno}

\newcommand{\cntext}[1]{\begin{CJK}{UTF8}{gbsn}#1\end{CJK}}
\renewcommand*\maketitle{%
  \thispagestyle{firstpage}
\begingroup
    \if@wideboxfn
    \setlength\bibindent{1.4\parindent}
    \else
    \setlength\bibindent{\parindent}
    \fi
    \renewcommand*\thefootnote{\@fnsymbol\c@footnote}%
    \renewcommand\@makefntext[1]{%
    \ifaa@longfn\hsize\textwidth\fi
    \noindent
    \hb@xt@\bibindent{\hss\@makefnmark\enspace}##1}
  \ifaa@twocolumn
  \begin{aa@strip}
    \aa@maketitle
    \@thanks
  \end{aa@strip}
  \else
    \begingroup
      \let\thanks\footnote
      \aa@maketitle
    \endgroup
  \fi
\endgroup
  \setcounter{footnote}{0}%
}
\makeatother
\usepackage{graphicx}
\usepackage{comment}
\usepackage[encapsulated]{CJK}
\usepackage{ucs}
\usepackage{txfonts}
\usepackage[breaklinks=true]{hyperref}
\newcommand{\orcid}[1]{\protect\href{https://orcid.org/#1}{\protect\includegraphics[width=8pt]{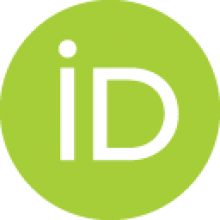}}}

\def\uv{$(u,v)$\xspace}

\def\m87{M87*}

\def\comrade{\texttt{Comrade}\xspace}
\def\doghit{\texttt{DoG-HIT}\xspace}
\def\moead{\texttt{MOEA/D}\xspace}
\def\ehtim{\texttt{ehtim}\xspace}
\def\difmap{\texttt{Difmap}\xspace}

\def\themis{\textsc{Themis}\xspace}

\makeatletter

\begin{document}
    \title{Probing jet base emission of \m87 with the 2021 \\ Event Horizon Telescope observations}
    \titlerunning{Jet base emission of \m87 in 2021 EHT Observations}
    \author{\tiny
            Saurabh\orcid{0000-0001-7156-4848}\inst{\ref{inst6}}\and
            Hendrik Müller\orcid{0000-0002-9250-0197}\inst{\ref{instn1}}\and
            Sebastiano D. von Fellenberg\orcid{0000-0002-9156-2249}\inst{\ref{inst131},\ref{inst6}}\and
            Paul Tiede\orcid{0000-0003-3826-5648}\inst{\ref{inst10},\ref{inst3}}\and
            Michael Janssen\orcid{0000-0001-8685-6544}\inst{\ref{inst29},\ref{inst6}}\and
            Lindy Blackburn\orcid{0000-0002-9030-642X}\inst{\ref{inst10},\ref{inst3}}\and
            Avery E. Broderick\orcid{0000-0002-3351-760X}\inst{\ref{inst26},\ref{inst27},\ref{inst28}}\and
            Erandi Chavez\inst{\ref{inst10}} \and
            Boris Georgiev\orcid{0000-0002-3586-6424}\inst{\ref{inst14}}\and
            Thomas P. Krichbaum\orcid{0000-0002-4892-9586}\inst{\ref{inst6}}\and
            Kotaro Moriyama\orcid{0000-0003-1364-3761}\inst{\ref{inst47},\ref{inst79}}\and
            Dhanya G. Nair\orcid{0000-0001-5357-7805}\inst{\ref{inst16},\ref{inst6}}\and
            Iniyan Natarajan\orcid{0000-0001-8242-4373}\inst{\ref{inst10},\ref{inst3}}\and
            Jongho Park\orcid{0000-0001-6558-9053}\inst{\ref{inst130},\ref{inst11}}\and
            Andrew Thomas West\inst{\ref{inst14}}\and
            Maciek Wielgus\orcid{0000-0002-8635-4242}\inst{\ref{inst5}}
            \\The Event Horizon Telescope Collaboration\\
            Kazunori Akiyama\orcid{0000-0002-9475-4254}\inst{\ref{inst1},\ref{inst2},\ref{inst3}}\and
            Ezequiel Albentosa-Ruíz\orcid{0000-0002-7816-6401}\inst{\ref{inst4}}\and
            Antxon Alberdi\orcid{0000-0002-9371-1033}\inst{\ref{inst5}}\and
            Walter Alef\inst{\ref{inst6}}\and
            Juan Carlos Algaba\orcid{0000-0001-6993-1696}\inst{\ref{inst7}}\and
            Richard Anantua\orcid{0000-0003-3457-7660}\inst{\ref{inst8},\ref{inst9},\ref{inst3},\ref{inst10}}\and
            Keiichi Asada\orcid{0000-0001-6988-8763}\inst{\ref{inst11}}\and
            Rebecca Azulay\orcid{0000-0002-2200-5393}\inst{\ref{inst4},\ref{inst12},\ref{inst6}}\and
            Uwe Bach\orcid{0000-0002-7722-8412}\inst{\ref{inst6}}\and
            Anne-Kathrin Baczko\orcid{0000-0003-3090-3975}\inst{\ref{inst13},\ref{inst6}}\and
            David Ball\inst{\ref{inst14}}\and
            Mislav Baloković\orcid{0000-0003-0476-6647}\inst{\ref{inst15}}\and
            Bidisha Bandyopadhyay\orcid{0000-0002-2138-8564}\inst{\ref{inst16}}\and
            John Barrett\orcid{0000-0002-9290-0764}\inst{\ref{inst1}}\and
            Michi Bauböck\orcid{0000-0002-5518-2812}\inst{\ref{inst17}}\and
            Bradford A. Benson\orcid{0000-0002-5108-6823}\inst{\ref{inst18},\ref{inst19}}\and
            Dan Bintley\inst{\ref{inst20},\ref{inst21}}\and
            Raymond Blundell\orcid{0000-0002-5929-5857}\inst{\ref{inst10}}\and
            Katherine L. Bouman\orcid{0000-0003-0077-4367}\inst{\ref{inst22}}\and
            Geoffrey C. Bower\orcid{0000-0003-4056-9982}\inst{\ref{inst20},\ref{inst21},\ref{inst23},\ref{inst24}}\and
            Michael Bremer\inst{\ref{inst25}}\and
            Roger Brissenden\orcid{0000-0002-2556-0894}\inst{\ref{inst10}}\and
            Silke Britzen\orcid{0000-0001-9240-6734}\inst{\ref{inst6}}\and
            Dominique Broguiere\orcid{0000-0001-9151-6683}\inst{\ref{inst25}}\and
            Thomas Bronzwaer\orcid{0000-0003-1151-3971}\inst{\ref{inst29}}\and
            Sandra Bustamante\orcid{0000-0001-6169-1894}\inst{\ref{inst30}}\and
            Douglas F. Carlos\orcid{0000-0002-1340-7702}\inst{\ref{inst31}}\and
            John E. Carlstrom\orcid{0000-0002-2044-7665}\inst{\ref{inst32},\ref{inst19},\ref{inst33},\ref{inst34}}\and
            Andrew Chael\orcid{0000-0003-2966-6220}\inst{\ref{inst35}}\and
            Chi-kwan Chan\orcid{0000-0001-6337-6126}\inst{\ref{inst14},\ref{inst36},\ref{inst37}}\and
            Dominic O. Chang\orcid{0000-0001-9939-5257}\inst{\ref{inst10},\ref{inst3}}\and
            Koushik Chatterjee\orcid{0000-0002-2825-3590}\inst{\ref{inst38},\ref{inst3},\ref{inst10}}\and
            Shami Chatterjee\orcid{0000-0002-2878-1502}\inst{\ref{inst39}}\and
            Ming-Tang Chen\orcid{0000-0001-6573-3318}\inst{\ref{inst23}}\and
            Yongjun Chen (\cntext{陈永军})\orcid{0000-0001-5650-6770}\inst{\ref{inst40},\ref{inst41}}\and
            Xiaopeng Cheng\orcid{0000-0003-4407-9868}\inst{\ref{inst42}}\and
            Paul Chichura\orcid{0000-0002-5397-9035}\inst{\ref{inst33},\ref{inst32}}\and
            Ilje Cho\orcid{0000-0001-6083-7521}\inst{\ref{inst42},\ref{inst43},\ref{inst5}}\and
            Pierre Christian\orcid{0000-0001-6820-9941}\inst{\ref{inst44}}\and
            Nicholas S. Conroy\orcid{0000-0003-2886-2377}\inst{\ref{inst45},\ref{inst10}}\and
            John E. Conway\orcid{0000-0003-2448-9181}\inst{\ref{inst13}}\and
            Thomas M. Crawford\orcid{0000-0001-9000-5013}\inst{\ref{inst19},\ref{inst32}}\and
            Geoffrey B. Crew\orcid{0000-0002-2079-3189}\inst{\ref{inst1}}\and
            Alejandro Cruz-Osorio\orcid{0000-0002-3945-6342}\inst{\ref{inst46},\ref{inst47}}\and
            Yuzhu Cui (\cntext{崔玉竹})\orcid{0000-0001-6311-4345}\inst{\ref{inst48}}\and
            Brandon Curd\orcid{0000-0002-8650-0879}\inst{\ref{inst8},\ref{inst3},\ref{inst10}}\and
            Rohan Dahale\orcid{0000-0001-6982-9034}\inst{\ref{inst5}}\and
            Jordy Davelaar\orcid{0000-0002-2685-2434}\inst{\ref{inst49},\ref{inst50}}\and
            Mariafelicia De Laurentis\orcid{0000-0002-9945-682X}\inst{\ref{inst51},\ref{inst52}}\and
            Roger Deane\orcid{0000-0003-1027-5043}\inst{\ref{inst53},\ref{inst54},\ref{inst55}}\and
            Gregory Desvignes\orcid{0000-0003-3922-4055}\inst{\ref{inst6},\ref{inst56}}\and
            Jason Dexter\orcid{0000-0003-3903-0373}\inst{\ref{inst57}}\and
            Vedant Dhruv\orcid{0000-0001-6765-877X}\inst{\ref{inst17}}\and
            Indu K. Dihingia\orcid{0000-0002-4064-0446}\inst{\ref{inst58}}\and
            Sheperd S. Doeleman\orcid{0000-0002-9031-0904}\inst{\ref{inst10},\ref{inst3}}\and
            Sergio A. Dzib\orcid{0000-0001-6010-6200}\inst{\ref{inst6}}\and
            Ralph P. Eatough\orcid{0000-0001-6196-4135}\inst{\ref{inst59},\ref{inst6}}\and
            Razieh Emami\orcid{0000-0002-2791-5011}\inst{\ref{inst10}}\and
            Heino Falcke\orcid{0000-0002-2526-6724}\inst{\ref{inst29}}\and
            Joseph Farah\orcid{0000-0003-4914-5625}\inst{\ref{inst60},\ref{inst61}}\and
            Vincent L. Fish\orcid{0000-0002-7128-9345}\inst{\ref{inst1}}\and
            Edward Fomalont\orcid{0000-0002-9036-2747}\inst{\ref{inst62}}\and
            H. Alyson Ford\orcid{0000-0002-9797-0972}\inst{\ref{inst14}}\and
            Marianna Foschi\orcid{0000-0001-8147-4993}\inst{\ref{inst5}}\and
            Raquel Fraga-Encinas\orcid{0000-0002-5222-1361}\inst{\ref{inst29}}\and
            William T. Freeman\inst{\ref{inst63},\ref{inst64}}\and
            Per Friberg\orcid{0000-0002-8010-8454}\inst{\ref{inst20},\ref{inst21}}\and
            Christian M. Fromm\orcid{0000-0002-1827-1656}\inst{\ref{inst65},\ref{inst47},\ref{inst6}}\and
            Antonio Fuentes\orcid{0000-0002-8773-4933}\inst{\ref{inst5}}\and
            Peter Galison\orcid{0000-0002-6429-3872}\inst{\ref{inst3},\ref{inst66},\ref{inst67}}\and
            Charles F. Gammie\orcid{0000-0001-7451-8935}\inst{\ref{inst17},\ref{inst45},\ref{inst68}}\and
            Roberto García\orcid{0000-0002-6584-7443}\inst{\ref{inst25}}\and
            Olivier Gentaz\orcid{0000-0002-0115-4605}\inst{\ref{inst25}}\and
            Ciriaco Goddi\orcid{0000-0002-2542-7743}\inst{\ref{inst31},\ref{inst69},\ref{inst70},\ref{inst71}}\and
            Roman Gold\orcid{0000-0003-2492-1966}\inst{\ref{inst72},\ref{inst73},\ref{inst74}}\and
            Arturo I. Gómez-Ruiz\orcid{0000-0001-9395-1670}\inst{\ref{inst75},\ref{inst76}}\and
            José L. Gómez\orcid{0000-0003-4190-7613}\inst{\ref{inst5}}\and
            Minfeng Gu (\cntext{顾敏峰})\orcid{0000-0002-4455-6946}\inst{\ref{inst40},\ref{inst77}}\and
            Mark Gurwell\orcid{0000-0003-0685-3621}\inst{\ref{inst10}}\and
            Kazuhiro Hada\orcid{0000-0001-6906-772X}\inst{\ref{inst78},\ref{inst79}}\and
            Daryl Haggard\orcid{0000-0001-6803-2138}\inst{\ref{inst80},\ref{inst81}}\and
            Ronald Hesper\orcid{0000-0003-1918-6098}\inst{\ref{inst82}}\and
            Dirk Heumann\orcid{0000-0002-7671-0047}\inst{\ref{inst14}}\and
            Luis C. Ho (\cntext{何子山})\orcid{0000-0001-6947-5846}\inst{\ref{inst83},\ref{inst84}}\and
            Paul Ho\orcid{0000-0002-3412-4306}\inst{\ref{inst11},\ref{inst21},\ref{inst20}}\and
            Mareki Honma\orcid{0000-0003-4058-9000}\inst{\ref{inst79},\ref{inst85},\ref{inst86}}\and
            Chih-Wei L. Huang\orcid{0000-0001-5641-3953}\inst{\ref{inst11}}\and
            Lei Huang (\cntext{黄磊})\orcid{0000-0002-1923-227X}\inst{\ref{inst40},\ref{inst77}}\and
            David H. Hughes\inst{\ref{inst75}}\and
            Shiro Ikeda\orcid{0000-0002-2462-1448}\inst{\ref{inst2},\ref{inst87},\ref{inst88},\ref{inst89}}\and
            C. M. Violette Impellizzeri\orcid{0000-0002-3443-2472}\inst{\ref{inst90},\ref{inst62}}\and
            Makoto Inoue\orcid{0000-0001-5037-3989}\inst{\ref{inst11}}\and
            Sara Issaoun\orcid{0000-0002-5297-921X}\inst{\ref{inst10},\ref{inst50}}\and
            David J. James\orcid{0000-0001-5160-4486}\inst{\ref{inst91},\ref{inst92}}\and
            Buell T. Jannuzi\orcid{0000-0002-1578-6582}\inst{\ref{inst14}}\and
            Britton Jeter\orcid{0000-0003-2847-1712}\inst{\ref{inst11}}\and
            Wu Jiang (\cntext{江悟})\orcid{0000-0001-7369-3539}\inst{\ref{inst40}}\and
            Alejandra Jiménez-Rosales\orcid{0000-0002-2662-3754}\inst{\ref{inst29}}\and
            Michael D. Johnson\orcid{0000-0002-4120-3029}\inst{\ref{inst10},\ref{inst3}}\and
            Svetlana Jorstad\orcid{0000-0001-6158-1708}\inst{\ref{inst93}}\and
            Adam C. Jones\inst{\ref{inst19}}\and
            Abhishek V. Joshi\orcid{0000-0002-2514-5965}\inst{\ref{inst17}}\and
            Taehyun Jung\orcid{0000-0001-7003-8643}\inst{\ref{inst42},\ref{inst94}}\and
            Ramesh Karuppusamy\orcid{0000-0002-5307-2919}\inst{\ref{inst6}}\and
            Tomohisa Kawashima\orcid{0000-0001-8527-0496}\inst{\ref{inst95}}\and
            Garrett K. Keating\orcid{0000-0002-3490-146X}\inst{\ref{inst10}}\and
            Mark Kettenis\orcid{0000-0002-6156-5617}\inst{\ref{inst96}}\and
            Dong-Jin Kim\orcid{0000-0002-7038-2118}\inst{\ref{inst97}}\and
            Jae-Young Kim\orcid{0000-0001-8229-7183}\inst{\ref{inst98}}\and
            Jongsoo Kim\orcid{0000-0002-1229-0426}\inst{\ref{inst42}}\and
            Junhan Kim\orcid{0000-0002-4274-9373}\inst{\ref{inst99}}\and
            Motoki Kino\orcid{0000-0002-2709-7338}\inst{\ref{inst2},\ref{inst100}}\and
            Jun Yi Koay\orcid{0000-0002-7029-6658}\inst{\ref{inst101},\ref{inst11}}\and
            Prashant Kocherlakota\orcid{0000-0001-7386-7439}\inst{\ref{inst3},\ref{inst10}}\and
            Yutaro Kofuji\inst{\ref{inst79},\ref{inst86}}\and
            Patrick M. Koch\orcid{0000-0003-2777-5861}\inst{\ref{inst11}}\and
            Shoko Koyama\orcid{0000-0002-3723-3372}\inst{\ref{inst101},\ref{inst11}}\and
            Carsten Kramer\orcid{0000-0002-4908-4925}\inst{\ref{inst25}}\and
            Joana A. Kramer\orcid{0009-0003-3011-0454}\inst{\ref{inst6}}\and
            Michael Kramer\orcid{0000-0002-4175-2271}\inst{\ref{inst6}}\and
            Cheng-Yu Kuo\orcid{0000-0001-6211-5581}\inst{\ref{inst102},\ref{inst11}}\and
            Noemi La Bella\orcid{0000-0002-8116-9427}\inst{\ref{inst29}}\and
            Deokhyeong Lee\orcid{0009-0003-2122-9437}\inst{\ref{inst103}}\and
            Sang-Sung Lee\orcid{0000-0002-6269-594X}\inst{\ref{inst42}}\and
            Aviad Levis\orcid{0000-0001-7307-632X}\inst{\ref{inst22}}\and
            Shaoliang Li\orcid{0009-0005-0338-9490}\inst{\ref{inst20},\ref{inst21}}\and
            Zhiyuan Li (\cntext{李志远})\orcid{0000-0003-0355-6437}\inst{\ref{inst104},\ref{inst105}}\and
            Rocco Lico\orcid{0000-0001-7361-2460}\inst{\ref{inst106},\ref{inst5}}\and
            Greg Lindahl\orcid{0000-0002-6100-4772}\inst{\ref{inst107}}\and
            Michael Lindqvist\orcid{0000-0002-3669-0715}\inst{\ref{inst13}}\and
            Mikhail Lisakov\orcid{0000-0001-6088-3819}\inst{\ref{inst108}}\and
            Jun Liu (\cntext{刘俊})\orcid{0000-0002-7615-7499}\inst{\ref{inst6}}\and
            Kuo Liu\orcid{0000-0002-2953-7376}\inst{\ref{inst40},\ref{inst41}}\and
            Elisabetta Liuzzo\orcid{0000-0003-0995-5201}\inst{\ref{inst109}}\and
            Wen-Ping Lo\orcid{0000-0003-1869-2503}\inst{\ref{inst11},\ref{inst110}}\and
            Andrei P. Lobanov\orcid{0000-0003-1622-1484}\inst{\ref{inst6}}\and
            Laurent Loinard\orcid{0000-0002-5635-3345}\inst{\ref{inst111},\ref{inst3},\ref{inst112}}\and
            Colin J. Lonsdale\orcid{0000-0003-4062-4654}\inst{\ref{inst1}}\and
            Amy E. Lowitz\orcid{0000-0002-4747-4276}\inst{\ref{inst14}}\and
            Ru-Sen Lu (\cntext{路如森})\orcid{0000-0002-7692-7967}\inst{\ref{inst40},\ref{inst41},\ref{inst6}}\and
            Nicholas R. MacDonald\orcid{0000-0002-6684-8691}\inst{\ref{inst6}}\and
            Jirong Mao (\cntext{毛基荣})\orcid{0000-0002-7077-7195}\inst{\ref{inst113},\ref{inst114},\ref{inst115}}\and
            Nicola Marchili\orcid{0000-0002-5523-7588}\inst{\ref{inst109},\ref{inst6}}\and
            Sera Markoff\orcid{0000-0001-9564-0876}\inst{\ref{inst116},\ref{inst117}}\and
            Daniel P. Marrone\orcid{0000-0002-2367-1080}\inst{\ref{inst14}}\and
            Alan P. Marscher\orcid{0000-0001-7396-3332}\inst{\ref{inst93}}\and
            Iván Martí-Vidal\orcid{0000-0003-3708-9611}\inst{\ref{inst4},\ref{inst12}}\and
            Satoki Matsushita\orcid{0000-0002-2127-7880}\inst{\ref{inst11}}\and
            Lynn D. Matthews\orcid{0000-0002-3728-8082}\inst{\ref{inst1}}\and
            Lia Medeiros\orcid{0000-0003-2342-6728}\inst{\ref{inst118}}\and
            Karl M. Menten\orcid{0000-0001-6459-0669}\inst{\ref{inst6},\ref{inst119}}\thanks{Deceased} \and
            Hugo Messias\orcid{0000-0002-2985-7994}\inst{\ref{inst_u1},\ref{inst_u2}}\and
            Izumi Mizuno\orcid{0000-0002-7210-6264}\inst{\ref{inst20},\ref{inst21}}\and
            Yosuke Mizuno\orcid{0000-0002-8131-6730}\inst{\ref{inst58},\ref{inst120},\ref{inst47}}\and
            Joshua Montgomery\orcid{0000-0003-0345-8386}\inst{\ref{inst81},\ref{inst19}}\and
            James M. Moran\orcid{0000-0002-3882-4414}\inst{\ref{inst10},\ref{inst3}}\and
            Monika Moscibrodzka\orcid{0000-0002-4661-6332}\inst{\ref{inst29}}\and
            Wanga Mulaudzi\orcid{0000-0003-4514-625X}\inst{\ref{inst116}}\and
            Cornelia Müller\orcid{0000-0002-2739-2994}\inst{\ref{inst6},\ref{inst29}}\and
            Alejandro Mus\orcid{0000-0003-0329-6874}\inst{\ref{inst69},\ref{inst106},\ref{inst121},\ref{inst122}}\and
            Gibwa Musoke\orcid{0000-0003-1984-189X}\inst{\ref{inst116},\ref{inst29}}\and
            Ioannis Myserlis\orcid{0000-0003-3025-9497}\inst{\ref{inst123}}\and
            Hiroshi Nagai\orcid{0000-0003-0292-3645}\inst{\ref{inst2},\ref{inst85}}\and
            Neil M. Nagar\orcid{0000-0001-6920-662X}\inst{\ref{inst16}}\and
            Masanori Nakamura\orcid{0000-0001-6081-2420}\inst{\ref{inst124},\ref{inst11}}\and
            Gopal Narayanan\orcid{0000-0002-4723-6569}\inst{\ref{inst30}}\and
            Antonios Nathanail\orcid{0000-0002-1655-9912}\inst{\ref{inst125},\ref{inst47}}\and
            Santiago Navarro Fuentes\inst{\ref{inst123}}\and
            Joey Neilsen\orcid{0000-0002-8247-786X}\inst{\ref{inst126}}\and
            Chunchong Ni\orcid{0000-0003-1361-5699}\inst{\ref{inst27},\ref{inst28},\ref{inst26}}\and
            Michael A. Nowak\orcid{0000-0001-6923-1315}\inst{\ref{inst127}}\and
            Junghwan Oh\orcid{0000-0002-4991-9638}\inst{\ref{inst96}}\and
            Hiroki Okino\orcid{0000-0003-3779-2016}\inst{\ref{inst79},\ref{inst86}}\and
            Héctor Raúl Olivares Sánchez\orcid{0000-0001-6833-7580}\inst{\ref{inst128}}\and
            Tomoaki Oyama\orcid{0000-0003-4046-2923}\inst{\ref{inst79}}\and
            Feryal Özel\orcid{0000-0003-4413-1523}\inst{\ref{inst129}}\and
            Daniel C. M. Palumbo\orcid{0000-0002-7179-3816}\inst{\ref{inst3},\ref{inst10}}\and
            Georgios Filippos Paraschos\orcid{0000-0001-6757-3098}\inst{\ref{inst6}}\and
            Harriet Parsons\orcid{0000-0002-6327-3423}\inst{\ref{inst20},\ref{inst21}}\and
            Nimesh Patel\orcid{0000-0002-6021-9421}\inst{\ref{inst10}}\and
            Ue-Li Pen\orcid{0000-0003-2155-9578}\inst{\ref{inst11},\ref{inst26},\ref{inst131},\ref{inst132},\ref{inst133}}\and
            Dominic W. Pesce\orcid{0000-0002-5278-9221}\inst{\ref{inst10},\ref{inst3}}\and
            Vincent Piétu\inst{\ref{inst25}}\and
            Alexander Plavin\orcid{0000-0003-2914-8554}\inst{\ref{inst3},\ref{inst10},\ref{inst6}}\and
            Aleksandar PopStefanija\inst{\ref{inst30}}\and
            Oliver Porth\orcid{0000-0002-4584-2557}\inst{\ref{inst116},\ref{inst47}}\and
            Ben Prather\orcid{0000-0002-0393-7734}\inst{\ref{inst17}}\and
            Giacomo Principe\orcid{0000-0003-0406-7387}\inst{\ref{inst134},\ref{inst135},\ref{inst106}}\and
            Dimitrios Psaltis\orcid{0000-0003-1035-3240}\inst{\ref{inst129}}\and
            Hung-Yi Pu\orcid{0000-0001-9270-8812}\inst{\ref{inst136},\ref{inst137},\ref{inst11}}\and
            Alexandra Rahlin\orcid{0000-0003-3953-1776}\inst{\ref{inst19}}\and
            Venkatessh Ramakrishnan\orcid{0000-0002-9248-086X}\inst{\ref{inst16},\ref{inst138},\ref{inst139}}\and
            Ramprasad Rao\orcid{0000-0002-1407-7944}\inst{\ref{inst10}}\and
            Mark G. Rawlings\orcid{0000-0002-6529-202X}\inst{\ref{inst140},\ref{inst20},\ref{inst21}}\and
            Luciano Rezzolla\orcid{0000-0002-1330-7103}\inst{\ref{inst47},\ref{inst141},\ref{inst142}}\and
            Angelo Ricarte\orcid{0000-0001-5287-0452}\inst{\ref{inst3},\ref{inst10}}\and
            Luca Ricci\orcid{0000-0002-4175-3194}\inst{\ref{inst143}}\and
            Bart Ripperda\orcid{0000-0002-7301-3908}\inst{\ref{inst131},\ref{inst144},\ref{inst132},\ref{inst26}}\and
            Jan Röder\orcid{0000-0002-2426-927X}\inst{\ref{inst5}}\and
            Freek Roelofs\orcid{0000-0001-5461-3687}\inst{\ref{inst29}}\and
            Cristina Romero-Cañizales\orcid{0000-0001-6301-9073}\inst{\ref{inst11}}\and
            Eduardo Ros\orcid{0000-0001-9503-4892}\inst{\ref{inst6}}\and
            Arash Roshanineshat\orcid{0000-0002-8280-9238}\inst{\ref{inst14}}\and
            Helge Rottmann\inst{\ref{inst6}}\and
            Alan L. Roy\orcid{0000-0002-1931-0135}\inst{\ref{inst6}}\and
            Ignacio Ruiz\orcid{0000-0002-0965-5463}\inst{\ref{inst123}}\and
            Chet Ruszczyk\orcid{0000-0001-7278-9707}\inst{\ref{inst1}}\and
            Kazi L. J. Rygl\orcid{0000-0003-4146-9043}\inst{\ref{inst109}}\and
            León D. S. Salas\orcid{0000-0003-1979-6363}\inst{\ref{inst116}}\and
            Salvador Sánchez\orcid{0000-0002-8042-5951}\inst{\ref{inst123}}\and
            David Sánchez-Argüelles\orcid{0000-0002-7344-9920}\inst{\ref{inst75},\ref{inst76}}\and
            Miguel Sánchez-Portal\orcid{0000-0003-0981-9664}\inst{\ref{inst123}}\and
            Mahito Sasada\orcid{0000-0001-5946-9960}\inst{\ref{inst145},\ref{inst79},\ref{inst146}}\and
            Kaushik Satapathy\orcid{0000-0003-0433-3585}\inst{\ref{inst14}}\and
            Tuomas Savolainen\orcid{0000-0001-6214-1085}\inst{\ref{inst147},\ref{inst139},\ref{inst6}}\and
            F. Peter Schloerb\inst{\ref{inst30}}\and
            Jonathan Schonfeld\orcid{0000-0002-8909-2401}\inst{\ref{inst10}}\and
            Karl-Friedrich Schuster\orcid{0000-0003-2890-9454}\inst{\ref{inst25}}\and
            Lijing Shao\orcid{0000-0002-1334-8853}\inst{\ref{inst84},\ref{inst6}}\and
            Zhiqiang Shen (\cntext{沈志强})\orcid{0000-0003-3540-8746}\inst{\ref{inst40},\ref{inst41}}\and
            Sasikumar Silpa\orcid{0000-0003-0667-7074}\inst{\ref{inst16}}\and
            Des Small\orcid{0000-0003-3723-5404}\inst{\ref{inst96}}\and
            Randall Smith\orcid{0000-0003-4284-4167}\inst{\ref{inst10}}\and
            Bong Won Sohn\orcid{0000-0002-4148-8378}\inst{\ref{inst42},\ref{inst94},\ref{inst43}}\and
            Jason SooHoo\orcid{0000-0003-1938-0720}\inst{\ref{inst1}}\and
            Kamal Souccar\orcid{0000-0001-7915-5272}\inst{\ref{inst30}}\and
            Joshua S. Stanway\orcid{0009-0003-7659-4642}\inst{\ref{inst148}}\and
            He Sun (\cntext{孙赫})\orcid{0000-0003-1526-6787}\inst{\ref{inst149},\ref{inst150}}\and
            Fumie Tazaki\orcid{0000-0003-0236-0600}\inst{\ref{inst151}}\and
            Alexandra J. Tetarenko\orcid{0000-0003-3906-4354}\inst{\ref{inst152}}\and
            Remo P. J. Tilanus\orcid{0000-0002-6514-553X}\inst{\ref{inst14},\ref{inst29},\ref{inst90},\ref{inst153}}\and
            Michael Titus\orcid{0000-0001-9001-3275}\inst{\ref{inst1}}\and
            Kenji Toma\orcid{0000-0002-7114-6010}\inst{\ref{inst154},\ref{inst155}}\and
            Pablo Torne\orcid{0000-0001-8700-6058}\inst{\ref{inst123},\ref{inst6}}\and
            Teresa Toscano\orcid{0000-0003-3658-7862}\inst{\ref{inst5}}\and
            Efthalia Traianou\orcid{0000-0002-1209-6500}\inst{\ref{inst5},\ref{inst6}}\and
            Tyler Trent\inst{\ref{inst14}}\and
            Sascha Trippe\orcid{0000-0003-0465-1559}\inst{\ref{inst156},\ref{inst157}}\and
            Matthew Turk\orcid{0000-0002-5294-0198}\inst{\ref{inst45}}\and
            Ilse van Bemmel\orcid{0000-0001-5473-2950}\inst{\ref{inst158}}\and
            Huib Jan van Langevelde\orcid{0000-0002-0230-5946}\inst{\ref{inst96},\ref{inst90},\ref{inst159}}\and
            Daniel R. van Rossum\orcid{0000-0001-7772-6131}\inst{\ref{inst29}}\and
            Jesse Vos\orcid{0000-0003-3349-7394}\inst{\ref{inst160}}\and
            Jan Wagner\orcid{0000-0003-1105-6109}\inst{\ref{inst6}}\and
            Derek Ward-Thompson\orcid{0000-0003-1140-2761}\inst{\ref{inst148}}\and
            John Wardle\orcid{0000-0002-8960-2942}\inst{\ref{inst161}}\and
            Jasmin E. Washington\orcid{0000-0002-7046-0470}\inst{\ref{inst14}}\and
            Jonathan Weintroub\orcid{0000-0002-4603-5204}\inst{\ref{inst10},\ref{inst3}}\and
            Robert Wharton\orcid{0000-0002-7416-5209}\inst{\ref{inst6}}\and
            Kaj Wiik\orcid{0000-0002-0862-3398}\inst{\ref{inst162},\ref{inst138},\ref{inst139}}\and
            Gunther Witzel\orcid{0000-0003-2618-797X}\inst{\ref{inst6}}\and
            Michael F. Wondrak\orcid{0000-0002-6894-1072}\inst{\ref{inst29},\ref{inst163}}\and
            George N. Wong\orcid{0000-0001-6952-2147}\inst{\ref{inst164},\ref{inst35}}\and
            Jompoj Wongphexhauxsorn\orcid{0000-0002-7730-4956}\inst{\ref{inst143},\ref{inst6}}\and
            Qingwen Wu (\cntext{吴庆文})\orcid{0000-0003-4773-4987}\inst{\ref{inst165}}\and
            Nitika Yadlapalli\orcid{0000-0003-3255-4617}\inst{\ref{inst22}}\and
            Paul Yamaguchi\orcid{0000-0002-6017-8199}\inst{\ref{inst10}}\and
            Aristomenis Yfantis\orcid{0000-0002-3244-7072}\inst{\ref{inst29}}\and
            Doosoo Yoon\orcid{0000-0001-8694-8166}\inst{\ref{inst116}}\and
            André Young\orcid{0000-0003-0000-2682}\inst{\ref{inst29}}\and
            Ziri Younsi\orcid{0000-0001-9283-1191}\inst{\ref{inst166},\ref{inst47}}\and
            Wei Yu (\cntext{于威})\orcid{0000-0002-5168-6052}\inst{\ref{inst10}}\and
            Feng Yuan (\cntext{袁峰})\orcid{0000-0003-3564-6437}\inst{\ref{inst167}}\and
            Ye-Fei Yuan (\cntext{袁业飞})\orcid{0000-0002-7330-4756}\inst{\ref{inst168}}\and
            Ai-Ling Zeng (\cntext{曾艾玲})\orcid{0009-0000-9427-4608}\inst{\ref{inst5}}\and
            J. Anton Zensus\orcid{0000-0001-7470-3321}\inst{\ref{inst6}}\and
            Shuo Zhang\orcid{0000-0002-2967-790X}\inst{\ref{inst169}}\and
            Guang-Yao Zhao\orcid{0000-0002-4417-1659}\inst{\ref{inst6},\ref{inst5}}\and
            Shan-Shan Zhao (\cntext{赵杉杉})\orcid{0000-0002-9774-3606}\inst{\ref{inst40}}
            }
    \institute{
            Max-Planck-Institut für Radioastronomie, Auf dem Hügel 69, D-53121 Bonn, Germany\label{inst6}\and
            Jansky Fellow of National Radio Astronomy Observatory, 1011 Lopezville Rd, Socorro, NM 87801, USA\label{instn1}\and
            Canadian Institute for Theoretical Astrophysics, University of Toronto, 60 St. George Street, Toronto, ON M5S 3H8, Canada\label{inst131}\and
            Center for Astrophysics $|$ Harvard \& Smithsonian, 60 Garden Street, Cambridge, MA 02138, USA\label{inst10}\and
            Black Hole Initiative at Harvard University, 20 Garden Street, Cambridge, MA 02138, USA\label{inst3}\and
            Department of Astrophysics, Institute for Mathematics, Astrophysics and Particle Physics (IMAPP), Radboud University, P.O. Box 9010, 6500 GL Nijmegen, The Netherlands\label{inst29}\and
            Perimeter Institute for Theoretical Physics, 31 Caroline Street North, Waterloo, ON N2L 2Y5, Canada\label{inst26}\and
            Department of Physics and Astronomy, University of Waterloo, 200 University Avenue West, Waterloo, ON N2L 3G1, Canada\label{inst27}\and
            Waterloo Centre for Astrophysics, University of Waterloo, Waterloo, ON N2L 3G1, Canada\label{inst28}\and
            Steward Observatory and Department of Astronomy, University of Arizona, 933 N. Cherry Ave., Tucson, AZ 85721, USA\label{inst14}\and
            Institut für Theoretische Physik, Goethe-Universität Frankfurt, Max-von-Laue-Straße 1, D-60438 Frankfurt am Main, Germany\label{inst47}\and
            Mizusawa VLBI Observatory, National Astronomical Observatory of Japan, 2-12 Hoshigaoka, Mizusawa, Oshu, Iwate 023-0861, Japan\label{inst79}\and
            School of Space Research, Kyung Hee University, 1732, Deogyeong-daero, Giheung-gu, Yongin-si, Gyeonggi-do 17104, Republic of Korea\label{inst130}\and
            Institute of Astronomy and Astrophysics, Academia Sinica, 11F of Astronomy-Mathematics Building, AS/NTU No. 1, Sec. 4, Roosevelt Rd., Taipei 106216, Taiwan, R.O.C.\label{inst11}\and
            Instituto de Astrofísica de Andalucía-CSIC, Glorieta de la Astronomía s/n, E-18008 Granada, Spain\label{inst5}\and
            Massachusetts Institute of Technology Haystack Observatory, 99 Millstone Road, Westford, MA 01886, USA\label{inst1}\and
            National Astronomical Observatory of Japan, 2-21-1 Osawa, Mitaka, Tokyo 181-8588, Japan\label{inst2}\and
            Departament d'Astronomia i Astrofísica, Universitat de València, C. Dr. Moliner 50, E-46100 Burjassot, València, Spain\label{inst4}\and
            Department of Physics, Faculty of Science, Universiti Malaya, 50603 Kuala Lumpur, Malaysia\label{inst7}\and
            Department of Physics \& Astronomy, The University of Texas at San Antonio, One UTSA Circle, San Antonio, TX 78249, USA\label{inst8}\and
            Physics \& Astronomy Department, Rice University, Houston, TX 77005-1827, USA\label{inst9}\and
            Observatori Astronòmic, Universitat de València, C. Catedrático José Beltrán 2, E-46980 Paterna, València, Spain\label{inst12}\and
            Department of Space, Earth and Environment, Chalmers University of Technology, Onsala Space Observatory, SE-43992 Onsala, Sweden\label{inst13}\and
            Yale Center for Astronomy \& Astrophysics, Yale University, 52 Hillhouse Avenue, New Haven, CT 06511, USA\label{inst15}\and
            Astronomy Department, Universidad de Concepción, Casilla 160-C, Concepción, Chile\label{inst16}\and
            Department of Physics, University of Illinois, 1110 West Green Street, Urbana, IL 61801, USA\label{inst17}\and
            Fermi National Accelerator Laboratory, MS209, P.O. Box 500, Batavia, IL 60510, USA\label{inst18}\and
            Department of Astronomy and Astrophysics, University of Chicago, 5640 South Ellis Avenue, Chicago, IL 60637, USA\label{inst19}\and
            East Asian Observatory, 660 N. A'ohoku Place, Hilo, HI 96720, USA\label{inst20}\and
            James Clerk Maxwell Telescope (JCMT), 660 N. A'ohoku Place, Hilo, HI 96720, USA\label{inst21}\and
            California Institute of Technology, 1200 East California Boulevard, Pasadena, CA 91125, USA\label{inst22}\and
            Institute of Astronomy and Astrophysics, Academia Sinica, 645 N. A'ohoku Place, Hilo, HI 96720, USA\label{inst23}\and
            Department of Physics and Astronomy, University of Hawaii at Manoa, 2505 Correa Road, Honolulu, HI 96822, USA\label{inst24}\and
            Institut de Radioastronomie Millimétrique (IRAM), 300 rue de la Piscine, F-38406 Saint Martin d'Hères, France\label{inst25}\and
            Department of Astronomy, University of Massachusetts, Amherst, MA 01003, USA\label{inst30}\and
            Instituto de Astronomia, Geofísica e Ciências Atmosféricas, Universidade de São Paulo, R. do Matão, 1226, São Paulo, SP 05508-090, Brazil\label{inst31}\and
            Kavli Institute for Cosmological Physics, University of Chicago, 5640 South Ellis Avenue, Chicago, IL 60637, USA\label{inst32}\and
            Department of Physics, University of Chicago, 5720 South Ellis Avenue, Chicago, IL 60637, USA\label{inst33}\and
            Enrico Fermi Institute, University of Chicago, 5640 South Ellis Avenue, Chicago, IL 60637, USA\label{inst34}\and
            Princeton Gravity Initiative, Jadwin Hall, Princeton University, Princeton, NJ 08544, USA\label{inst35}\and
            Data Science Institute, University of Arizona, 1230 N. Cherry Ave., Tucson, AZ 85721, USA\label{inst36}\and
            Program in Applied Mathematics, University of Arizona, 617 N. Santa Rita, Tucson, AZ 85721, USA\label{inst37}\and
            Department of Physics, University of Maryland, 7901 Regents Drive, College Park, MD 20742, USA\label{inst38}\and
            Cornell Center for Astrophysics and Planetary Science, Cornell University, Ithaca, NY 14853, USA\label{inst39}\and
            Shanghai Astronomical Observatory, Chinese Academy of Sciences, 80 Nandan Road, Shanghai 200030, People's Republic of China\label{inst40}\and
            Key Laboratory of Radio Astronomy and Technology, Chinese Academy of Sciences, A20 Datun Road, Chaoyang District, Beijing, 100101, People’s Republic of China\label{inst41}\and
            Korea Astronomy and Space Science Institute, Daedeok-daero 776, Yuseong-gu, Daejeon 34055, Republic of Korea\label{inst42}\and
            Department of Astronomy, Yonsei University, Yonsei-ro 50, Seodaemun-gu, 03722 Seoul, Republic of Korea\label{inst43}\and
            WattTime, 490 43rd Street, Unit 221, Oakland, CA 94609, USA\label{inst44}\and
            Department of Astronomy, University of Illinois at Urbana-Champaign, 1002 West Green Street, Urbana, IL 61801, USA\label{inst45}\and
            Instituto de Astronomía, Universidad Nacional Autónoma de México (UNAM), Apdo Postal 70-264, Ciudad de México, México\label{inst46}\and
            Institute of Astrophysics, Central China Normal University, Wuhan 430079, People's Republic of China\label{inst48}\and
            Department of Astrophysical Sciences, Peyton Hall, Princeton University, Princeton, NJ 08544, USA\label{inst49}\and
            NASA Hubble Fellowship Program, Einstein Fellow\label{inst50}\and
            Dipartimento di Fisica ``E. Pancini'', Università di Napoli ``Federico II'', Compl. Univ. di Monte S. Angelo, Edificio G, Via Cinthia, I-80126, Napoli, Italy\label{inst51}\and
            INFN Sez. di Napoli, Compl. Univ. di Monte S. Angelo, Edificio G, Via Cinthia, I-80126, Napoli, Italy\label{inst52}\and
            Wits Centre for Astrophysics, University of the Witwatersrand, 1 Jan Smuts Avenue, Braamfontein, Johannesburg 2050, South Africa\label{inst53}\and
            Department of Physics, University of Pretoria, Hatfield, Pretoria 0028, South Africa\label{inst54}\and
            Centre for Radio Astronomy Techniques and Technologies, Department of Physics and Electronics, Rhodes University, Makhanda 6140, South Africa\label{inst55}\and
            LESIA, Observatoire de Paris, Université PSL, CNRS, Sorbonne Université, Université de Paris, 5 place Jules Janssen, F-92195 Meudon, France\label{inst56}\and
            JILA and Department of Astrophysical and Planetary Sciences, University of Colorado, Boulder, CO 80309, USA\label{inst57}\and
            Tsung-Dao Lee Institute, Shanghai Jiao Tong University, Shengrong Road 520, Shanghai, 201210, People’s Republic of China\label{inst58}\and
            National Astronomical Observatories, Chinese Academy of Sciences, 20A Datun Road, Chaoyang District, Beijing 100101, PR China\label{inst59}\and
            Las Cumbres Observatory, 6740 Cortona Drive, Suite 102, Goleta, CA 93117-5575, USA\label{inst60}\and
            Department of Physics, University of California, Santa Barbara, CA 93106-9530, USA\label{inst61}\and
            National Radio Astronomy Observatory, 520 Edgemont Road, Charlottesville, VA 22903, USA\label{inst62}\and
            Department of Electrical Engineering and Computer Science, Massachusetts Institute of Technology, 32-D476, 77 Massachusetts Ave., Cambridge, MA 02142, USA\label{inst63}\and
            Google Research, 355 Main St., Cambridge, MA 02142, USA\label{inst64}\and
            Institut für Theoretische Physik und Astrophysik, Universität Würzburg, Emil-Fischer-Str. 31, D-97074 Würzburg, Germany\label{inst65}\and
            Department of History of Science, Harvard University, Cambridge, MA 02138, USA\label{inst66}\and
            Department of Physics, Harvard University, Cambridge, MA 02138, USA\label{inst67}\and
            NCSA, University of Illinois, 1205 W. Clark St., Urbana, IL 61801, USA\label{inst68}\and
            Dipartimento di Fisica, Università degli Studi di Cagliari, SP Monserrato-Sestu km 0.7, I-09042 Monserrato (CA), Italy\label{inst69}\and
            INAF - Osservatorio Astronomico di Cagliari, via della Scienza 5, I-09047 Selargius (CA), Italy\label{inst70}\and
            INFN, sezione di Cagliari, I-09042 Monserrato (CA), Italy\label{inst71}\and
            Institute for Mathematics and Interdisciplinary Center for Scientific Computing, Heidelberg University, Im Neuenheimer Feld 205, Heidelberg 69120, Germany\label{inst72}\and
            Institut f\"ur Theoretische Physik, Universit\"at Heidelberg, Philosophenweg 16, 69120 Heidelberg, Germany\label{inst73}\and
            CP3-Origins, University of Southern Denmark, Campusvej 55, DK-5230 Odense, Denmark\label{inst74}\and
            Instituto Nacional de Astrofísica, Óptica y Electrónica. Apartado Postal 51 y 216, 72000. Puebla Pue., México\label{inst75}\and
            Consejo Nacional de Humanidades, Ciencia y Tecnología, Av. Insurgentes Sur 1582, 03940, Ciudad de México, México\label{inst76}\and
            Key Laboratory for Research in Galaxies and Cosmology, Chinese Academy of Sciences, Shanghai 200030, People's Republic of China\label{inst77}\and
            Graduate School of Science, Nagoya City University, Yamanohata 1, Mizuho-cho, Mizuho-ku, Nagoya, 467-8501, Aichi, Japan\label{inst78}\and
            Department of Physics, McGill University, 3600 rue University, Montréal, QC H3A 2T8, Canada\label{inst80}\and
            Trottier Space Institute at McGill, 3550 rue University, Montréal,  QC H3A 2A7, Canada\label{inst81}\and
            NOVA Sub-mm Instrumentation Group, Kapteyn Astronomical Institute, University of Groningen, Landleven 12, 9747 AD Groningen, The Netherlands\label{inst82}\and
            Department of Astronomy, School of Physics, Peking University, Beijing 100871, People's Republic of China\label{inst83}\and
            Kavli Institute for Astronomy and Astrophysics, Peking University, Beijing 100871, People's Republic of China\label{inst84}\and
            Department of Astronomical Science, The Graduate University for Advanced Studies (SOKENDAI), 2-21-1 Osawa, Mitaka, Tokyo 181-8588, Japan\label{inst85}\and
            Department of Astronomy, Graduate School of Science, The University of Tokyo, 7-3-1 Hongo, Bunkyo-ku, Tokyo 113-0033, Japan\label{inst86}\and
            The Institute of Statistical Mathematics, 10-3 Midori-cho, Tachikawa, Tokyo, 190-8562, Japan\label{inst87}\and
            Department of Statistical Science, The Graduate University for Advanced Studies (SOKENDAI), 10-3 Midori-cho, Tachikawa, Tokyo 190-8562, Japan\label{inst88}\and
            Kavli Institute for the Physics and Mathematics of the Universe, The University of Tokyo, 5-1-5 Kashiwanoha, Kashiwa, 277-8583, Japan\label{inst89}\and
            Leiden Observatory, Leiden University, Postbus 2300, 9513 RA Leiden, The Netherlands\label{inst90}\and
            ASTRAVEO LLC, PO Box 1668, Gloucester, MA 01931, USA\label{inst91}\and
            Applied Materials Inc., 35 Dory Road, Gloucester, MA 01930, USA\label{inst92}\and
            Institute for Astrophysical Research, Boston University, 725 Commonwealth Ave., Boston, MA 02215, USA\label{inst93}\and
            University of Science and Technology, Gajeong-ro 217, Yuseong-gu, Daejeon 34113, Republic of Korea\label{inst94}\and
            National Institute of Technology, Ichinoseki College, Takanashi, Hagisho, Ichinoseki, Iwate, 021-8511, Japan\label{inst95}\and
            Joint Institute for VLBI ERIC (JIVE), Oude Hoogeveensedijk 4, 7991 PD Dwingeloo, The Netherlands\label{inst96}\and
            CSIRO, Space and Astronomy, PO Box 76, Epping, NSW 1710, Australia\label{inst97}\and
            Department of Physics, Ulsan National Institute of Science and Technology (UNIST), Ulsan 44919, Republic of Korea\label{inst98}\and
            Department of Physics, Korea Advanced Institute of Science and Technology (KAIST), 291 Daehak-ro, Yuseong-gu, Daejeon 34141, Republic of Korea\label{inst99}\and
            Kogakuin University of Technology \& Engineering, Academic Support Center, 2665-1 Nakano, Hachioji, Tokyo 192-0015, Japan\label{inst100}\and
            Graduate School of Science and Technology, Niigata University, 8050 Ikarashi 2-no-cho, Nishi-ku, Niigata 950-2181, Japan\label{inst101}\and
            Physics Department, National Sun Yat-Sen University, No. 70, Lien-Hai Road, Kaosiung City 80424, Taiwan, R.O.C.\label{inst102}\and
            Department of Astronomy, Kyungpook National University, 80 Daehak-ro, Buk-gu, Daegu 41566, Republic of Korea\label{inst103}\and
            School of Astronomy and Space Science, Nanjing University, Nanjing 210023, People's Republic of China\label{inst104}\and
            Key Laboratory of Modern Astronomy and Astrophysics, Nanjing University, Nanjing 210023, People's Republic of China\label{inst105}\and
            INAF-Istituto di Radioastronomia, Via P. Gobetti 101, I-40129 Bologna, Italy\label{inst106}\and
            Common Crawl Foundation, 9663 Santa Monica Blvd. 425, Beverly Hills, CA 90210 USA\label{inst107}\and
            Instituto de Física, Pontificia Universidad Católica de Valparaíso, Casilla 4059, Valparaíso, Chile\label{inst108}\and
            INAF-Istituto di Radioastronomia \& Italian ALMA Regional Centre, Via P. Gobetti 101, I-40129 Bologna, Italy\label{inst109}\and
            Department of Physics, National Taiwan University, No. 1, Sec. 4, Roosevelt Rd., Taipei 106216, Taiwan, R.O.C\label{inst110}\and
            Instituto de Radioastronomía y Astrofísica, Universidad Nacional Autónoma de México, Morelia 58089, México\label{inst111}\and
            David Rockefeller Center for Latin American Studies, Harvard University, 1730 Cambridge Street, Cambridge, MA 02138, USA\label{inst112}\and
            Yunnan Observatories, Chinese Academy of Sciences, 650011 Kunming, Yunnan Province, People's Republic of China\label{inst113}\and
            Center for Astronomical Mega-Science, Chinese Academy of Sciences, 20A Datun Road, Chaoyang District, Beijing, 100012, People's Republic of China\label{inst114}\and
            Key Laboratory for the Structure and Evolution of Celestial Objects, Chinese Academy of Sciences, 650011 Kunming, People's Republic of China\label{inst115}\and
            Anton Pannekoek Institute for Astronomy, University of Amsterdam, Science Park 904, 1098 XH, Amsterdam, The Netherlands\label{inst116}\and
            Gravitation and Astroparticle Physics Amsterdam (GRAPPA) Institute, University of Amsterdam, Science Park 904, 1098 XH Amsterdam, The Netherlands\label{inst117}\and
            Center for Gravitation, Cosmology and Astrophysics, Department of Physics, University of Wisconsin–Milwaukee, P.O. Box 413, Milwaukee, WI 53201, USA\label{inst118}\and
            Deceased\label{inst119}\and
            Joint ALMA Observatory, Alonso de C\'ordova 3107, Vitacura 763-0355, Santiago, Chile\label{inst_u1}\and
            European Southern Observatory, Alonso de C\'ordova 3107, Vitacura, Casilla 19001, Santiago, Chile\label{inst_u2}\and
            School of Physics and Astronomy, Shanghai Jiao Tong University, 800 Dongchuan Road, Shanghai, 200240, People’s Republic of China\label{inst120}\and
            SCOPIA Research Group, University of the Balearic Islands, Dept. of Mathematics and Computer Science, Ctra. Valldemossa, Km 7.5, Palma 07122, Spain\label{inst121}\and
            Artificial Intelligence Research Institute of the Balearic Islands (IAIB), Palma 07122, Spain\label{inst122}\and
            Institut de Radioastronomie Millimétrique (IRAM), Avenida Divina Pastora 7, Local 20, E-18012, Granada, Spain\label{inst123}\and
            National Institute of Technology, Hachinohe College, 16-1 Uwanotai, Tamonoki, Hachinohe City, Aomori 039-1192, Japan\label{inst124}\and
            Research Center for Astronomy, Academy of Athens, Soranou Efessiou 4, 115 27 Athens, Greece\label{inst125}\and
            Department of Physics, Villanova University, 800 Lancaster Avenue, Villanova, PA 19085, USA\label{inst126}\and
            Physics Department, Washington University, CB 1105, St. Louis, MO 63130, USA\label{inst127}\and
            Departamento de Matemática da Universidade de Aveiro and Centre for Research and Development in Mathematics and Applications (CIDMA), Campus de Santiago, 3810-193 Aveiro, Portugal\label{inst128}\and
            School of Physics, Georgia Institute of Technology, 837 State St NW, Atlanta, GA 30332, USA\label{inst129}\and
            Dunlap Institute for Astronomy and Astrophysics, University of Toronto, 50 St. George Street, Toronto, ON M5S 3H4, Canada\label{inst132}\and
            Canadian Institute for Advanced Research, 180 Dundas St West, Toronto, ON M5G 1Z8, Canada\label{inst133}\and
            Dipartimento di Fisica, Università di Trieste, I-34127 Trieste, Italy\label{inst134}\and
            INFN Sez. di Trieste, I-34127 Trieste, Italy\label{inst135}\and
            Department of Physics, National Taiwan Normal University, No. 88, Sec. 4, Tingzhou Rd., Taipei 116, Taiwan, R.O.C.\label{inst136}\and
            Center of Astronomy and Gravitation, National Taiwan Normal University, No. 88, Sec. 4, Tingzhou Road, Taipei 116, Taiwan, R.O.C.\label{inst137}\and
            Finnish Centre for Astronomy with ESO, University of Turku, FI-20014 Turun Yliopisto, Finland\label{inst138}\and
            Aalto University Metsähovi Radio Observatory, Metsähovintie 114, FI-02540 Kylmälä, Finland\label{inst139}\and
            Gemini Observatory/NSF NOIRLab, 670 N. A’ohōkū Place, Hilo, HI 96720, USA\label{inst140}\and
            Frankfurt Institute for Advanced Studies, Ruth-Moufang-Strasse 1, D-60438 Frankfurt, Germany\label{inst141}\and
            School of Mathematics, Trinity College, Dublin 2, Ireland\label{inst142}\and
            Julius-Maximilians-Universität Würzburg, Fakultät für Physik und Astronomie, Institut für Theoretische Physik und Astrophysik, Lehrstuhl für Astronomie, Emil-Fischer-Str. 31, D-97074 Würzburg, Germany\label{inst143}\and
            Department of Physics, University of Toronto, 60 St. George Street, Toronto, ON M5S 1A7, Canada\label{inst144}\and
            Department of Physics, Tokyo Institute of Technology, 2-12-1 Ookayama, Meguro-ku, Tokyo 152-8551, Japan\label{inst145}\and
            Hiroshima Astrophysical Science Center, Hiroshima University, 1-3-1 Kagamiyama, Higashi-Hiroshima, Hiroshima 739-8526, Japan\label{inst146}\and
            Aalto University Department of Electronics and Nanoengineering, PL 15500, FI-00076 Aalto, Finland\label{inst147}\and
            Jeremiah Horrocks Institute, University of Central Lancashire, Preston PR1 2HE, UK\label{inst148}\and
            National Biomedical Imaging Center, Peking University, Beijing 100871, People’s Republic of China\label{inst149}\and
            College of Future Technology, Peking University, Beijing 100871, People’s Republic of China\label{inst150}\and
            Tokyo Electron Technology Solutions Limited, 52 Matsunagane, Iwayado, Esashi, Oshu, Iwate 023-1101, Japan\label{inst151}\and
            Department of Physics and Astronomy, University of Lethbridge, Lethbridge, Alberta T1K 3M4, Canada\label{inst152}\and
            Netherlands Organisation for Scientific Research (NWO), Postbus 93138, 2509 AC Den Haag, The Netherlands\label{inst153}\and
            Frontier Research Institute for Interdisciplinary Sciences, Tohoku University, Sendai 980-8578, Japan\label{inst154}\and
            Astronomical Institute, Tohoku University, Sendai 980-8578, Japan\label{inst155}\and
            Department of Physics and Astronomy, Seoul National University, Gwanak-gu, Seoul 08826, Republic of Korea\label{inst156}\and
            SNU Astronomy Research Center, Seoul National University, Gwanak-gu, Seoul 08826, Republic of Korea\label{inst157}\and
            ASTRON, Oude Hoogeveensedijk 4, 7991 PD Dwingeloo, The Netherlands\label{inst158}\and
            University of New Mexico, Department of Physics and Astronomy, Albuquerque, NM 87131, USA\label{inst159}\and
            Centre for Mathematical Plasma Astrophysics, Department of Mathematics, KU Leuven, Celestijnenlaan 200B, B-3001 Leuven, Belgium\label{inst160}\and
            Physics Department, Brandeis University, 415 South Street, Waltham, MA 02453, USA\label{inst161}\and
            Tuorla Observatory, Department of Physics and Astronomy, University of Turku, FI-20014 Turun Yliopisto, Finland\label{inst162}\and
            Radboud Excellence Fellow of Radboud University, Nijmegen, The Netherlands\label{inst163}\and
            School of Natural Sciences, Institute for Advanced Study, 1 Einstein Drive, Princeton, NJ 08540, USA\label{inst164}\and
            School of Physics, Huazhong University of Science and Technology, Wuhan, Hubei, 430074, People's Republic of China\label{inst165}\and
            Mullard Space Science Laboratory, University College London, Holmbury St. Mary, Dorking, Surrey, RH5 6NT, UK\label{inst166}\and
            Center for Astronomy and Astrophysics and Department of Physics, Fudan University, Shanghai 200438, People's Republic of China\label{inst167}\and
            Astronomy Department, University of Science and Technology of China, Hefei 230026, People's Republic of China\label{inst168}\and
            Department of Physics and Astronomy, Michigan State University, 567 Wilson Rd, East Lansing, MI 48824, USA\label{inst169}
            }
            
\abstract{
        We investigate the presence and spatial characteristics of the jet base emission in M87* at $230~\mathrm{GHz}$, enabled by the significantly enhanced \uv coverage in the 2021 Event Horizon Telescope (EHT) observations. The integration of the $12\mathrm{-m}$ Kitt Peak Telescope (USA) and NOEMA (France) stations into the array introduces two critical intermediate-length baselines to SMT (USA) and IRAM 30$\mathrm{-m}$ (Spain), providing sensitivity to emission structures at spatial scales of $\sim 250~\mathrm{\mu as}$ and $\sim 2500~\mathrm{\mu as}$ ($\sim 0.02~\rm{pc}$ and $\sim 0.2~\rm{pc}$). Without these new baselines, previous EHT observations of the source in 2017 and 2018 lacked the capability to constrain emission on large scales, where a ``missing flux" of order $\sim 1\,$Jy is expected to reside.
        To probe these scales, we analyzed closure phases---robust against station-based gain calibration errors---and model the jet base emission using a simple Gaussian component offset from the compact ring emission at spatial separations $> 100~\mathrm{\mu as}$. 
        Our analysis revealed a Gaussian feature centered at ($\Delta \rm{R.A.} \approx 320~\mathrm{\mu as}$, $\Delta \rm{Dec.} \approx 60~\mathrm{\mu as}$), projected separation of $\approx 5500~\rm{AU}$, with an estimated flux density of only $\sim 60~\mathrm{mJy}$, implying that most of the missing flux identified in previous EHT studies had to originate from different, larger scales. Brighter emission at the relevant spatial scales is firmly ruled out, and the data do not favor more complex models.
        This component aligns with the inferred position of the large-scale jet and is therefore physically consistent with the emission of the jet base.
        While our findings point to detectable jet base emission at $230~\mathrm{GHz}$, the limited coverage provided by only two intermediate baselines limits our ability to robustly reconstruct its morphology. Consequently, we treated the recovered Gaussian as an upper limit on the jet base flux density. Future EHT observations with expanded intermediate baseline coverage will be essential to constrain the structure and nature of this component with higher precision.
    }
    
\bigskip
\maketitle
\section{Introduction}
\label{sec:intro}
\begin{figure*}
    \centering
    \includegraphics[width=\linewidth]{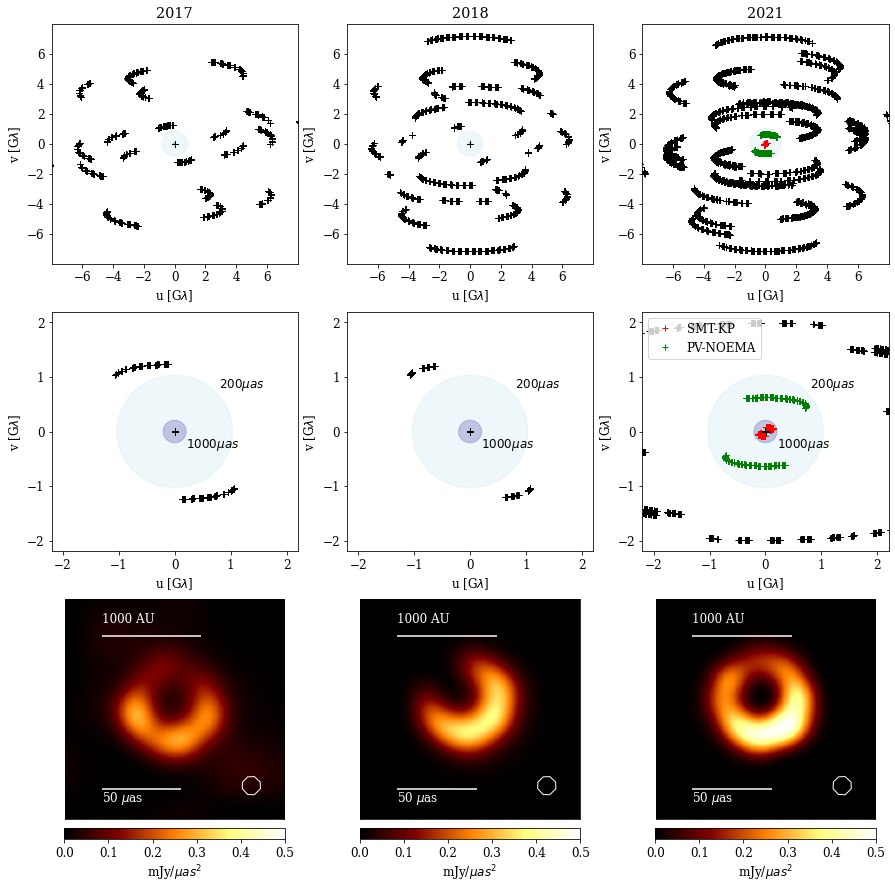}
    \caption{\uv coverage of the EHT and representative ring image obtained by \doghit \citep{M87_2021} in 2017, 2018, and 2021. Upper panels: \uv with the SMT-KP (red) and PV-NOEMA baselines (green) highlighted. These are used to constrain the jet base emission. Middle panels: a zoomed-in view of the \uv coverage, highlighting the short and intermediate baselines. We also highlight the \uv coordinates related to spatial scales of $1000\,\mu\mathrm{as}$ and $200\,\mu\mathrm{as}$, corresponding to the expected range for extended emission. Bottom panels: \doghit image of the central ring obtained from these data \citep{M87_2021}.}
    \label{fig: uv-coverage}
\end{figure*}

The giant elliptical galaxy Messier 87 (M87) serves as a cornerstone for understanding active galactic nuclei (AGNe), primarily due to its prominent relativistic jet powered by a supermassive black hole (SMBH) \m87 at its center. This large-scale jet is visible across the electromagnetic spectrum, including the infrared \citep[][]{Roder:2025ifv}, optical \citep[][]{Perlman2011}, X-ray \citep[][]{Marshall2002}, and gamma-ray regime \citep[][]{Abramowski2012}. In the radio regime, the jet in M87 and its dynamics have been monitored at multiple wavelengths and spatial scales \citep[e.g.][]{Walker2018, Lister2018, Kim2018, Kim2023, Cui2023, Lu2023}.
At milli-arcsecond scales, the jet displays hints of triple-peaked helical dynamics \citep{Asada2016, Hada2017, Nikonov2023}, likely connected to precession \citep{Cui2023}.
The envelope of the jet limb shows a quasi-parabolic profile over a wide range of spatial scales \citep{Asada2012, Hada2013, Nakamura2013, Nakamura2018, Walker2018, Lister2018}, which appears to persist down to just a few gravitational radii, as observed in the very long baseline interferometry (VLBI) observations from the Global Millimetre VLBI Array (GMVA) \citep{Kim2018, Lu2023, Kim2024} and space VLBI \citep{Kim2023}.

The Event Horizon Telescope (EHT) captured the first resolved image of the M87 nucleus at event horizon scales \citep{M87p1, M87p2, M87p3, M87p4, M87p5, M87p6}, revealing a ring with a central brightness depression interpreted as the shadow of the SMBH. Subsequent EHT studies examined this feature in linearly and circularly polarized light \citep{M87p7,M87p8,M87p9} and found the ring to be persistent between 2017 and 2018 \citep{M87_2018p1}. \cite{Wielgus2020} used model-fitting on proto-EHT data at 230\,GHz to study the ring evolution in the years prior to the first resolved images.

A recent polarimetric analysis of EHT data spanning three epochs---2017, 2018, and 2021---studied the dynamics across years of \m87 \citep{M87_2021}. It was found that the ring stays remarkably stable in total intensity (i.e., the ring diameter and thickness), with varying brightness asymmetry over the years; but the linear polarization fractions and polarization patterns vary drastically between the years. Although the milliarcsecond-scale jet has been studied extensively over the years \citep{Walker2018, Cui2023}, its direct connection to the jet base and the inner accretion flow is not well constrained \citep[e.g.,][]{Hada2024}. A major step forward was made through $86\,{\rm GHz}$ GMVA imaging \citep{Lu2023}, which revealed a ring-like structure with northern and southern components linking to a collimated, edge-brightened jet base. However, a central ridgeline present in the images may be an artifact from the deconvolution procedure \citep{Kim2024}. In this work, we aim to build on these findings and constrain the resolved, extended emission associated with the jet base at 230\,GHz using the 2021 EHT observations.

In general, interferometric observations can only constrain spatial scales corresponding to baseline lengths present in a given array \citep[e.g.,][]{Thompson2017}. Very long baselines (thousands of kilometers) resolve the finest angular scales, such as the $\sim40~\mu{\rm as}$ ring. So-called trivial intra-site baselines (a few hundred meters) are sensitive to arcsecond-scale structure but provide no resolution near the black hole \citep{Georgiev2025}. Intermediate-length or short baselines (a few hundred to a few thousand kilometers) probe angular scales of hundreds to thousands of microarcseconds, making them particularly important for studying the connection between the accretion flow and the jet base. 

While the presence of the large-scale jet at 230\,GHz is evident in Atacama Large Millimeter/submillimeter Array (ALMA) observations \citep{Goddi2021}, detecting and imaging this emission with VLBI at 230\,GHz is challenging for 2017 and 2018 EHT datasets lacking short or intermediate length baselines. This challenge results in a missing flux---flux density on the order of $\sim1.0 - 1.4~\mathrm{Jy}$ detected on trivial baselines from co-located stations (ALMA/APEX, JCMT/SMA)---that compact ring-only models (flux density $\sim0.5-1.0~\mathrm{Jy}$ depending on the imaging approach) cannot fully explain \citep{M87p4, M87_2018p1}. Although a robust detection was impossible due to the lack of short and intermediate length baselines, some hints of extended emission, particularly to the southwest of the ring, in the pre-2021 EHT data were reported in \cite{Broderick2022} and \cite{M87_2018p1} and through independent inspections of the residual maps in \cite{Carilli2022} and \cite{Arras2022}.

The EHTC's data analysis methodology applied various tools to deal with the missing flux issue, validated across datasets \citep{M87p4, M87_2018p1, M87_2021}. All methods have been explicitly validated using synthetic data generated with the 2021 coverage\footnote{That is, images of a variety of simple geometric structures, physically motivated structures, as well as simulations of accretion flow \citep{M87_2021}.}, in order to assess any potential systematic biases in the reconstructions. We refer to the synthetic data tests in \citet{M87_2021} for more details.

The regularized maximum likelihood (RML) methods typically deal with the missing flux problem by rescaling the flux density at intra-site baseline spacings. \doghit attempts to recover the image from closure quantities only (in arbitrary units) and scales the whole image structure globally to a flux density that minimizes the $\chi^2$ of the amplitudes. CLEAN methods add a large-scale Gaussian to account for the missing large-scale flux density. \comrade and \themis apply similar strategies, but typically with more degrees of freedom for the large-scale component. The Hybrid-\themis framework \citep{Broderick2020b} incorporates the model-fitting of a narrow ring, allowing more emission to be placed as jet base emission \citep{Broderick2022}.

The 2021 EHT observations offer the best $(u, v)$-coverage to date, providing a new opportunity to detect the faint, extended emission associated with the jet. In particular, this is aided by the addition of two new stations; the NOrthern Extended Millimeter Array (NOEMA) in France and the Kitt Peak 12-m telescope (KP) in the USA. These stations add intermediate baselines of $\sim1100\,{\rm km}$ and $\sim 100\,{\rm km}$ to the IRAM 30-m telescope on Pico Veleta (PV) in Spain and the Submillimeter Telescope (SMT) in the USA, respectively. The corresponding angular scales of these two baseline pairs are $\sim250~\mathrm{\mu as}$ and $\sim2500~\mathrm{\mu as}$, i.e., these baselines are in principle sensitive to extended emission that could not be detected by the EHT in previous years. However, each baseline only probes a narrow range of spatial frequencies. Due to the limited number of baselines sensitive to intermediate angular scales, the \uv coverage is too sparse to directly image the faint, extended structure. It would require several intermediate baselines with resolving power of a few hundred $\mathrm{\mu as}$ to robustly image the jet base region.

As a consequence, much like in the case of other very sparse VLBI data sets \citep[e.g.,][]{Wielgus2020}, we resort to a geometric model-fitting approach to constrain the presence of extended emission in the 2021 EHT data. For this purpose, we used closure phases, which are the sum of the visibility phases on a closed loop of three baselines (a triangle). This quantity is powerful because it is immune to station-dependent atmospheric and instrumental phase errors, making it a robust observable in VLBI \citep{Thompson2017, 2020ApJ...894...31B}. Therefore, we use closure triangles of the baselines of interest, formed with the most sensitive station in the array, ALMA. On the relatively short baselines that make up these triangles, the compact $\approx 40\,\mu{\rm as}$ ring is only marginally resolved. While these baselines individually have limited resolving power for such small-scale structures, closure phases remain sensitive to asymmetries in the overall brightness distribution. For a point source or a symmetric ring, the closure phases on these triangles are expected to be zero. However, when we compare the observed closure phases with those predicted by image reconstructions, we find non-zero residuals. This indicates that the observed phase structure cannot be fully explained by the ring alone--- even when intrinsic asymmetries in the ring are included---and instead points to the presence of additional, extended and asymmetric emission on angular scales probed by the intermediate baselines.

To test this interpretation, we model the extended emission using an offset Gaussian component in addition to the compact ring. Varying the position of the Gaussian component, we found that the most plausible location for this excess emission is to the South-West of the ring (i.e., the direction of the large-scale jet), consistent with the conclusions reached by, for e.g., \cite{Broderick2022}.

The analysis presented in this manuscript builds on the success of previous EHT studies, explicitly fitting a large-scale component, and does not question the validity of the recovered compact-scale images. As a complement to the compact-scale studies, we present a deeper analysis of emission features of the jet base at intermediate spatial scales ($\sim 250-2500~\mathrm{\mu as}$).

\section{EHT observations of \m87 in 2021}

In this section, we summarize the main data properties and results, with details provided in the companion paper \citep{M87_2021}. The data were correlated with \textsc{DiFX} \citep{Deller2007,Deller2011} using the output bands mode, a spectral window width of $58~\mathrm{MHz}$, and a bandwidth of $1856~\mathrm{MHz}$ per band. For this analysis, we used band $3$ and band $4$ data, centered on $227.1~\mathrm{GHz}$ and $228.1~\mathrm{GHz}$, respectively. The data were converted from linear to circular polarization feed basis using \textsc{PolConvert} \citep{MartiVidal2016}. Fringe-fitting was done using r\textsc{PICARD} \citep{Janssen2019, 2022Janssen, vonFellenberg2025}. 
The 2021 EHT observing campaign (April $9-19$) consisted of \m87 observations on April 9, 13, 14, 17 and 18. Since April 14 and 17 were relatively shorter tracks and NOEMA did not participate in April 13, we used only April 18. In 2021, NOEMA and KP joined the array for the first time, providing intermediate-length baselines that may be sensitive to the extended jet base emission (see Figure~\ref{fig: uv-coverage}).

In \citet{M87_2021}, the data were analyzed by seven teams using different imaging algorithms. The robustness of the reconstruction was demonstrated via cross-validation. Two teams used \difmap \citep{Shepherd_1997}, in combination with the AIPS task LPCAL \citep{Leppanen1995} and GPCAL \citep{Park2021b, Park2023a, Park2023b}. The other teams used the RML methods \doghit \citep{Mueller2022, Mueller2023a, Mueller2023b}, \ehtim \citep{Chael2016, Chael2018}, and \moead \citep{Mueller2023c, Mus2024a, Mus2024b}, as well as the Bayesian imaging techniques \themis \citep{Broderick2020} and \comrade \citep{Tiede2022}. Although these imaging algorithms rely on different assumptions, they converge on a consistent recovered image structure on event horizon scales \citep{M87_2021}. 
We show a representative image from this analysis obtained by \doghit in the bottom panels of Figure~\ref{fig: uv-coverage}. The 2021 observations of \m87 are well represented by a ring structure with an asymmetry oriented to the southwest. Although the total-intensity images are very consistent across years and methods, there is some significant evolution of the polarized structure over the years \citep[for more details, we refer the reader to][]{M87_2021}. In this work, we focused on total intensity and leave the discussion of polarization of extended components for future work.

\begin{figure}
    \includegraphics[width=\columnwidth]{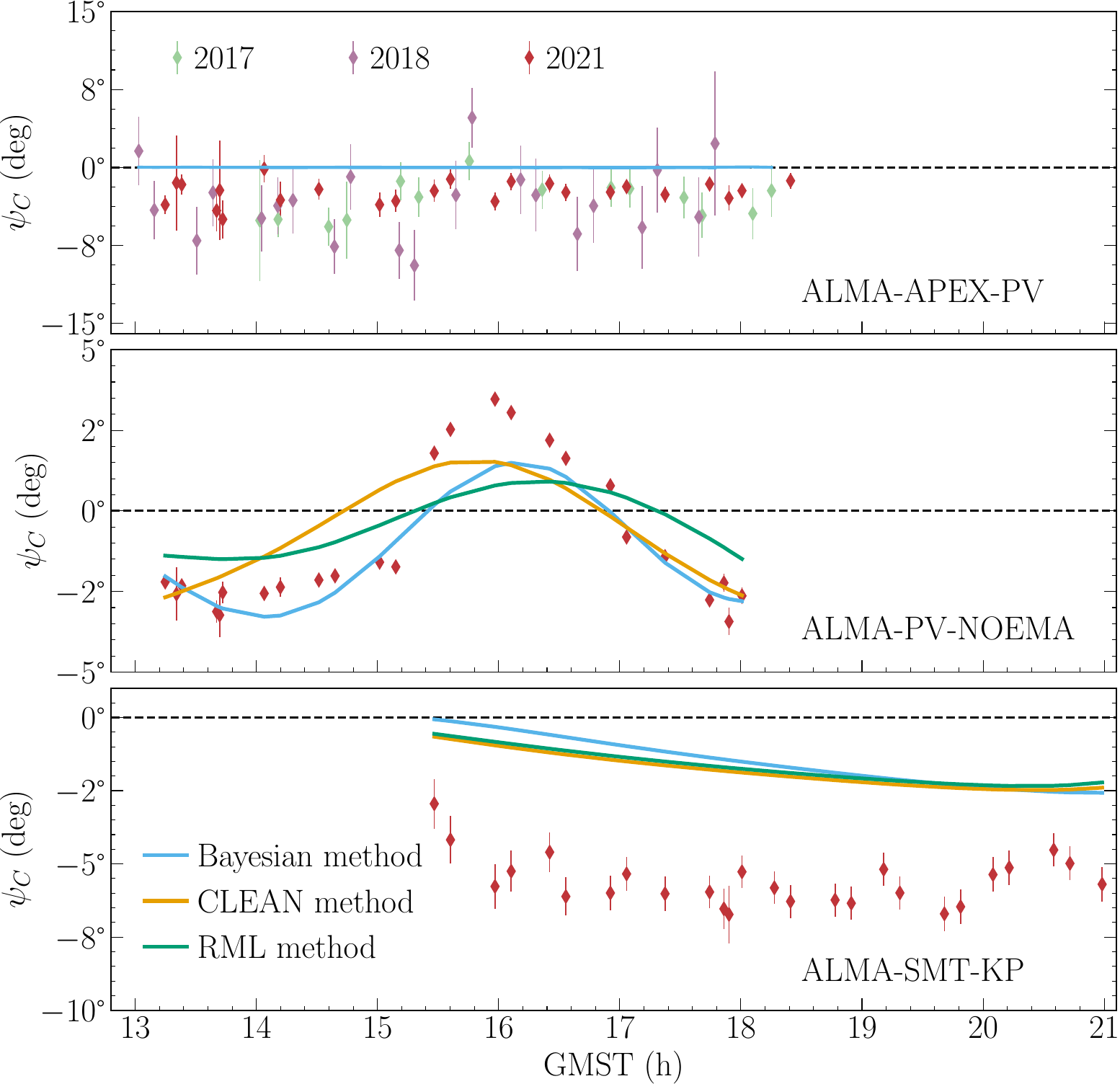}
    \caption{ALMA-APEX-PV (top) closure phases of \m87 as observed in 2017, 2018 and 2021. The ALMA-PV-NOEMA (middle) and ALMA-SMT-KP (bottom) closure phases on \m87 observed on April 18, 2021 (figure is adapted from \cite{M87_2021}) are shown together with ring-image models obtained with different imaging algorithms (see text for details). The top panel shows an example of a trivial closure phase triangle, where two stations are co-located, see \cite{Georgiev2025} for details. The image models are discrepant with the observed data by a few degrees, which may be attributable to intermediate-scale emission observed on the PV-NOEMA and SMT-KP baselines.}
    \label{fig:ring-2-data-comp}
\end{figure}
\begin{figure}
    \centering
    \includegraphics[width=\columnwidth]{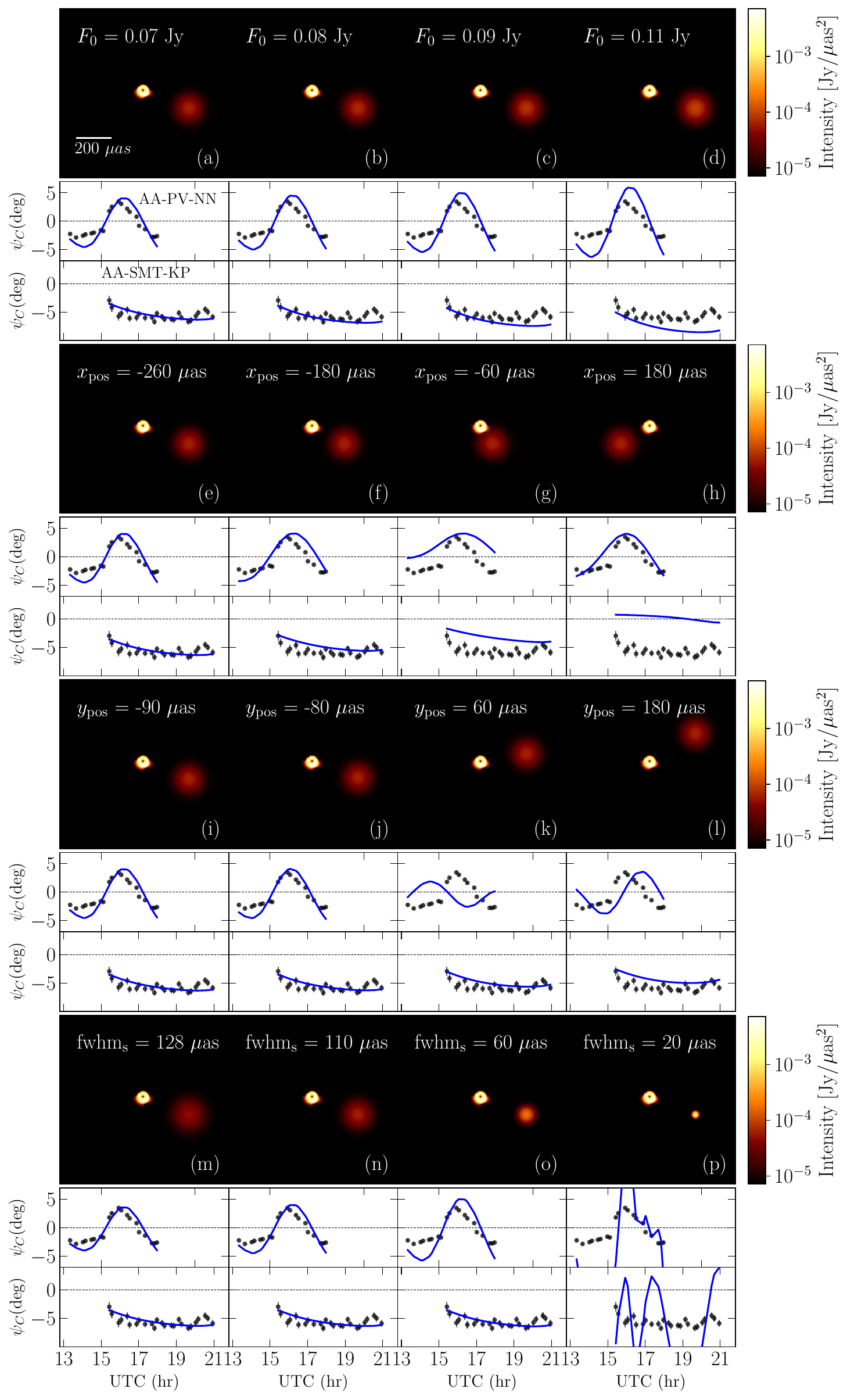}
    \caption{Symmetric Gaussian component with varying parameters over the reconstructed \doghit image as represented in Figure~\ref{fig: uv-coverage} represented in log scale. Different rows correspond to changes in the values of different parameters (see text for details). The solid lines represents the corresponding complete model (ring + Gaussian) closure phases on the respective triangles.}
    \label{fig: toy_gauss}
\end{figure}
\section{Jet base emission} \label{sec: non-ring-emission}

\begin{figure*}
    \centering
    \includegraphics[width=\textwidth]{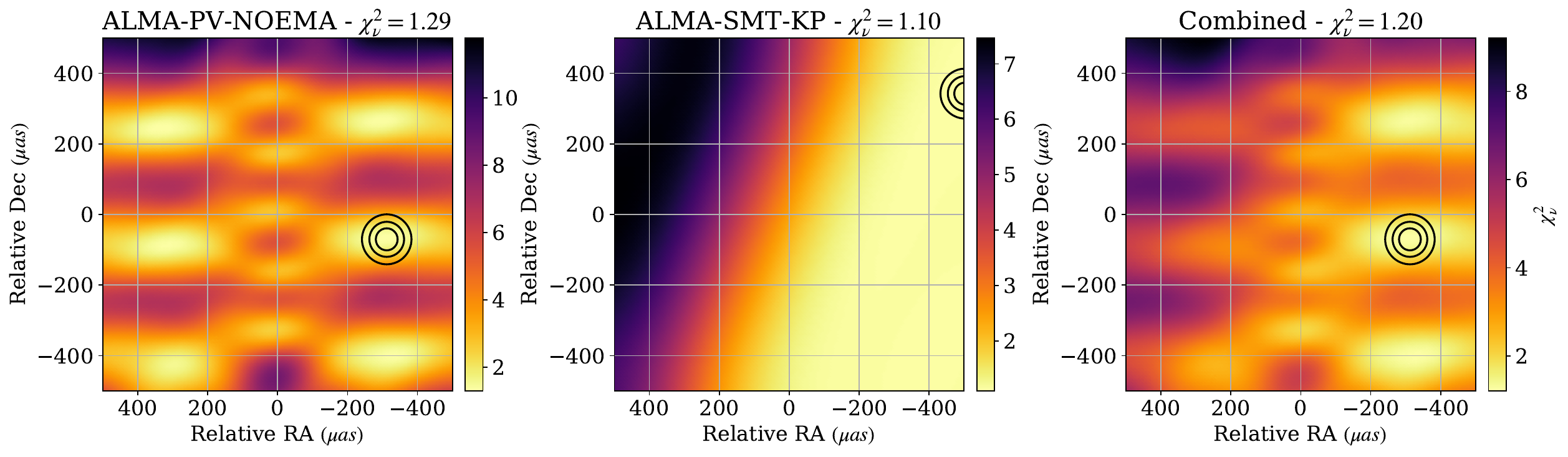}
    \caption{$\chi^2_{\nu}$ achieved with the Gaussian model for a fixed width and brightness ratio, but varying position. The left and middle panels show the $\chi^2_\nu$ for the ALMA-PV-NOEMA and the ALMA-SMT-KP triangles, respectively, and the right panel shows the combined $\chi^2_\nu$. The contours indicate the position of the best-fit model.}
    \label{fig: chisq_map}
\end{figure*}
\subsection{Bump and offset in closure phases}

We show the \uv coverage of the EHT in 2017, 2018, and 2021 in Figure~\ref{fig: uv-coverage} and highlight two new baselines introduced in 2021. The PV-NOEMA baseline is $\sim 1100~{\rm km}$ in length, corresponding to on-sky spatial scales $\sim250~\mathrm{\mu as}$; the SMT-KP baseline is $\sim100~{\rm km}$ in length, corresponding to $\sim2500~\mathrm{\mu as}$ scales. These intermediate baselines are sensitive to jet base emission, if it exists and is bright and compact enough to be detectable. However, each of these spatial scales is probed by a single baseline with a limited track length. Because of that, self-calibration techniques may inadvertently introduce or suppress extended emission if calibrated against a ring-only image or central point sources \citep[e.g.,][]{Carilli2022}. To overcome this problem, we analyze data that were not self-calibrated, and use closure phases that are insensitive to station-based gain errors \citep[e.g.,][]{2020ApJ...894...31B} to probe the jet base emission at scales accessible with SMT-KP and PV-NOEMA.

Three stations are required to construct closure phases. Because no other intermediate-length baselines are available, the closure triangles formed necessarily also probe small-scale structures. However, we can choose triangles that are particularly useful, i.e., show low variations during the track (indicating that the long baseline contribution is small or canceled), are long in duration, and have high signal-to-noise ratio (S/N). The high-sensitivity station ALMA satisfies all these requirements for both baselines of interest. We thus focused on the ALMA-SMT-KP and ALMA-PV-NOEMA triangles in this analysis.

Since the small-scale information of the long ALMA baselines in those triangles cannot be disentangled from the intermediate-scale information, the jet base information has to be derived from the difference between the data and the ring-only image. In Figure~\ref{fig:ring-2-data-comp} we show the closure phases on various triangles. The top panel displays an example triangle containing the ALMA-APEX baseline across the three years. The middle and bottom panels display the triangles of interest obtained for the small-scale ring images for different methods---one representative image is chosen for each framework---\difmap (CLEAN), \doghit (RML) and \comrade (Bayesian). None of the imaging methods (marked by solid lines) can fully recover the closure phases with the compact ring image alone on these triangles. However, we note that the different imaging algorithms applied and validated various strategies to mitigate the effect of jet base emission on the ring images. These include the addition of large-scale Gaussian components that are not used to create the fits in Figure~\ref{fig:ring-2-data-comp}. Figure~\ref{fig:ring-2-data-comp} displays the image fits without these additional factors, focusing solely on the compact ring feature.

In particular, the ALMA-PV-NOEMA triangle features a `bump' that is not well captured by the ring images, and, similarly, ALMA-SMT-KP has a visible `offset'. Due to the high sensitivity on those triangles, the difference is significant despite its small magnitude.

One plausible explanation for this residual of a few degrees in the closure phase is a faint but measurable emission on intermediate spatial scales sensed by PV-NOEMA and SMT-KP.
\subsection{Model} \label{sec: model}

In the following, we motivate a simple model to explain the observed closure phase residuals on the two selected triangles. The triangle that includes the longer baseline PV-NOEMA features a bump with a period of a few hours. Conversely, the ALMA-SMT-KP triangle does not have a beating but a simple offset from the ring-image models. This suggests that the relevant spatial scales may be on the order of, or slightly larger than, the PV-NOEMA scale, as any larger structure would introduce a beating on the ALMA-SMT-KP triangle as well. Moreover, it needs to be located close to the image center. Such a closure phase signature is characteristic of a ``binary model''; in our case it consists of the ring, which is not resolved by either of the two baselines, and an extra component.

Based on these simple considerations, we propose the following source model: 
the compact ring emission as seen on long baselines, with an additional, faint, Gaussian component located at scales $100-2000~\mathrm{\mu as}$ from the center of the ring emission. This model gives four additional degrees of freedom, the $x$ and $y$ positions, the brightness $F_0$, and the full width at half maximum ($\rm{fwhm}$) of the Gaussian component. The model is selected to be as simple as possible to restrict the number of degrees of freedom, which is a necessary assumption for the model fitting due to the limited information on jet base emission present in the data. Because the PV-NOEMA and SMT-KP baselines are sensitive to two different spatial scales, they are not necessarily described by the same model component. Thus the model's complexity can be trivially extended by allowing for an elliptical Gaussian component or even a second Gaussian component.

Deriving from these simple initial considerations, we can set approximate limits for the expected parameters of the additional Gaussian component. The size should be such that it is probed by the PV-NOEMA and SMT-KP baselines, creating a beating on the former within just two hours. Hence, it does not describe extended emission but most likely jet base emission with spatial scales similar to the ring diameter. 

We refer to detailed analysis of the total flux density of the ring feature in EHT observations presented in \citet{M87_2018p1}. Although for the arcsecond scale structure in \m87 typically flux densities between $1.0~\mathrm{Jy}$ and $2.0~\mathrm{Jy}$ were measured \citep{Bower2015,Wielgus2020}, for the milliarcsecond and microarcesond structure usually flux densities between $0.5$ and $1.0\,\mathrm{Jy}$ have been reported \citep{Doeleman2012, Akiyama2015, Kim2018, Wielgus2020, Lu2023}, consistent with the values found by the EHT \citep{M87p4, M87_2018p1, M87_2021}. The discrepancy between the ring image and the flux density detected with ALMA alone of about $1~\mathrm{Jy}$ is due to the existence of an (as of yet) undetectable large-scale jet. The additional Gaussian component fitted in this manuscript is likely to explain some, but not necessarily all, of the missing flux. This sets an approximate upper limit of $0.5~\mathrm{Jy}$ for the flux of the Gaussian component, preferably less.

\section{Localization of the emission region}

\subsection{Impact of the model parameters} \label{sec: impact_model}

In the following analysis, we examine the impact of jet base emission when modeled as a Gaussian component. We base our analysis on a \doghit reconstruction: since the algorithm separates small and large scale emission naturally, it is optimized to fit to the closure phases and closure amplitudes in the 2021 data. However, we note that the black hole shadow images obtained by various algorithms have similar characteristics on the ALMA-PV-NOEMA and ALMA-SMT-KP triangles with respect to the bump and the offset, see Figure~\ref{fig:ring-2-data-comp}.

\autoref{fig: toy_gauss} shows the ring image obtained by \doghit with an additional Gaussian added on top, varying some of the parameters described in Section~\ref{sec: model}. The upper part of the panels display the recovered central region with an additional Gaussian component in logarithmic scale. Panels e, f, h, and i show the observed closure phases (black markers) and model closure phases (solid lines) on the ALMA-PV-NOEMA triangle (AA-PV-NN) and ALMA-SMT-KP triangle (AA-MG-KP).

The addition of a Gaussian component creates a periodic modulation -- a beating -- on the PV-NOEMA baseline. There are locations (primarily to the east and west of the ring, see \autoref{fig: toy_gauss}) where the beating is broadly aligned with the data, and locations where it misaligns and worsens the fit to the closure phase (primarily to the north). On the ALMA-SMT-KP triangle, the component produces a closure phase offset, which matches the data for some locations (good fits primarily to the West). The brightness of the component scales the amplitude of the beating and offset on the respective triangles. An overly bright Gaussian component overestimates the beating and the correction of the offset, while a too dim component underestimates the necessary corrections (first row of Figure~\ref{fig: toy_gauss}). Finally, an overly compact component causes large-magnitude swings in the closure phase, while for a larger feature it is too smooth (fourth row in Figure~\ref{fig: toy_gauss}).

These findings are summarized in Figure~\ref{fig: chisq_map}. In the left panel we show an example of reduced $\chi^2$ ($\chi_{\nu}^2$) of the ALMA-PV-NOEMA triangle as a function of the location of the Gaussian component with a fixed flux density of $60\,\mathrm{mJy}$ and a size of $180\,\mu\mathrm{as}$. Locations $200\,\mathrm{\mu as}$ to the west or the east of the ring are preferred. 
Similarly, the middle panel shows the $\chi^2_\nu$ for the ALMA-SMT-KP triangle, with the preferred location to the southwest. 
In the right panel we show the combined $\chi^2_\nu$ and a preferred region to the southwest, roughly $300\,\mathrm{\mu as}$ to the West and $100\,\mathrm{\mu as}$ to the South of the ring. 
However, the presence of multiple regions with low $\chi^2_\nu$ indicates that the current model remains weakly constrained and degenerate.

\subsection{Symmetric Gaussian component} \label{sec: fiducial}

To obtain the maximum posterior position for a Gaussian component, we use the Markov chain Monte Carlo (MCMC) sampler \textsc{emcee} \citep{Foreman2013}.
We adopt uniform priors for the flux ($F_0 \in [0,0.2]\,\mathrm{Jy}$), the location ($x,y \in [-500,500]\,\mu\mathrm{as}$), and the width (${\rm fwhm_s }\in [50,250]\,\mu\mathrm{as}$). 
The maximum posterior parameters are given in Table~\ref{tab:params}.
The corner plots are shown in Appendix~\ref{appendix: corner plots} (\autoref{fig: cornerplot}). The best fit component has a flux density of $\sim 60~\mathrm{mJy}$ and is located towards the Southwestern portion of the ring.

\begin{figure*}
    \centering
    \includegraphics[width=\linewidth]{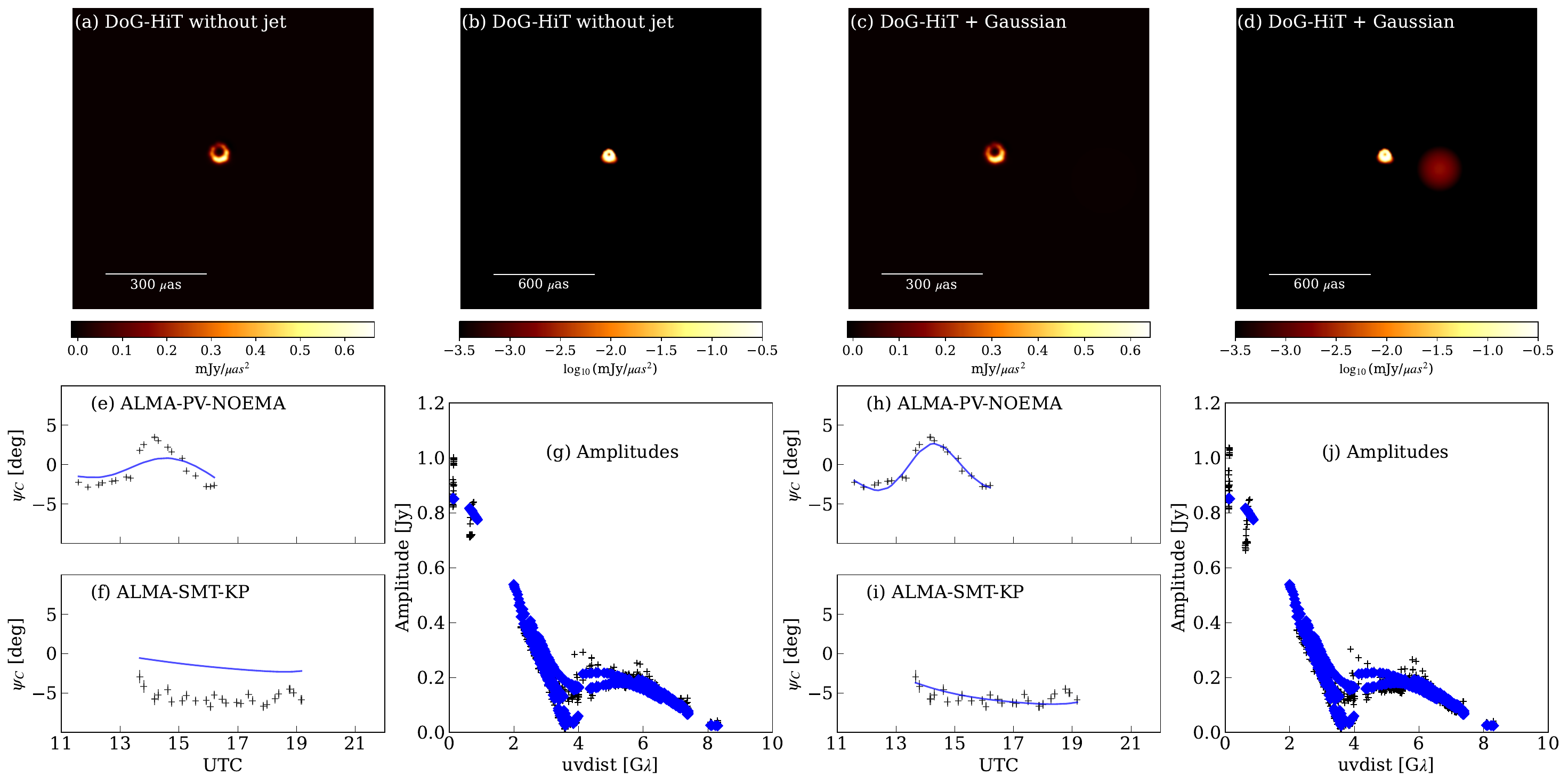}
    \caption{Ring-only \doghit image (panel a: linear scale, panel b: logarithmic scale) compared to the \doghit ring image with an additional Gaussian (panel c and d). In the bottom row, we show the fit to the bump and offset in closure phases and the fit to the amplitudes for \doghit (panels e,f,g) and for \doghit with the symmetric Gaussian component added (panels h,i,j).}
    \label{fig: fiducial_fit}
\end{figure*}

We show the obtained ring image as well as the ring image with an additional Gaussian component in Figure~\ref{fig: fiducial_fit}. The respective fits of the Gaussian components to the ALMA-PV-NOEMA and ALMA-SMT-KP closure phases are shown in the panels e, f, h, and i. We highlighted notable improvements in the fitting quality of the closure phases visible from panel h and i.
Despite the simplicity of the model, the $\chi_{\nu}^2$ is $1.29$ for the bump (compared to $2.73$ without the jet base emission), and $1.12$ for the offset (compared to $2.52$).
The component has almost negligible impact on all other triangles; the $\chi_{\nu}^2$ over all triangles improves from $1.5$ to $1.39$. The $\chi_{\nu}^2$ does not increase for any triangle, see Appendix~\ref{appendix: visbilities} for further details. In Figure~\ref{fig: amplitudes}, we show the impact of the additional Gaussian component on the amplitudes for different baselines. Amplitudes on baselines longer than $\sim2\,G\lambda$ stay mostly unaffected by the additional Gaussian component. This is reflected by the fit to the self-calibrated amplitudes shown in panel g and j in Figure~\ref{fig: fiducial_fit} and the left panel in Figure~\ref{fig: amplitudes}. The effect of the Gaussian component can be completely absorbed into the amplitude self-calibration.

Finally, synthetic data tests were performed in \citet{M87_2021}. These tests also include the validation of a ring image with an extended jet component. The results of this test for all seven imaging techniques are described in Appendix D3 of that paper. We reprint the main result of this validation in Appendix~\ref{appendix: visbilities}, Figure~\ref{fig: jet_test}. We note that all techniques were able to recover the ring emission correctly with exceptionally high cross-correlation $>0.995$ for the total-intensity 2021 data.

We conclude that the addition of the Gaussian model does not worsen the otherwise excellent fit to the data obtained in \citet{M87_2021} with the black hole shadow images only. Vice versa, the Gaussian component only leaves a significant effect on the ALMA-PV-NOEMA and ALMA-SMT-KP triangles, exhibiting a small effect on the overall fitting statistics. Hence, based on the success of the fitting in the aforementioned study, and the synthetic data tests that were performed in \citet{M87_2021}, we have no reason to question the validity of the ring emission presented in \citet{M87_2021}.

\begin{table}
    \centering
    \caption{Parameter estimates for the different Gaussian model fits described in \autoref{sec: fiducial}, \autoref{sec: asymmetry} and \autoref{sec: 2gaussian}.}
    \label{tab:params}
    \renewcommand{\arraystretch}{1.5}
    \begin{tabular}{l |r r r}
        \hline \hline
        Parameter & Symmetric  & Asymmetric  & Double \\ 
        \hline
        $F_0\,(\mathrm{mJy})$      & $70^{+10}_{-10}$& $60_{-20}^{+20}$ &  $50_{-30}^{+30}$\\
        $x_{1}\,(\mu\mathrm{as})$ & $-304_{-38}^{+38}$& $-300_{-37}^{+37}$     & $-297_{-46}^{+46}$\\
        $y_{1}\,(\mu\mathrm{as})$ & $-68_{-12}^{+12}$ & $-63_{-12}^{+12}$      & $-68_{-18}^{+18}$\\
        ${\rm fwhm_{1}} (\mu\mathrm{as})$ & $169_{-27}^{+27}$ & $180_{-35}^{+35}$ & $150_{-56}^{+56}$\\
        \hline
        $\mathcal{R}_1$           &$-$& $0.7_{-0.3}^{+0.3}$  & $0.6_{-0.5}^{+0.5}$\\
        $\Phi_1\,(^\circ)$        &$-$& $14_{-119}^{+119}$  & $23_{-101}^{+101}$\\
        \hline
        $F_2\,(\rm{mJy})$          &$-$& $-$                & $20_{-20}^{+20}$ \\
        $x_{2}\,(\mu\mathrm{as})$ &$-$& $-$ & $-291_{-96}^{+96}$ \\
        $y_{2}\,(\mu\mathrm{as})$ &$-$& $-$ & $117_{-126}^{+126}$ \\
        ${\rm fwhm_{2}}\,(\mu\mathrm{as})$& $-$& $-$& $150_{-56}^{+56}$ \\
        $\mathcal{R}_2$           &$-$& $-$ & $0.5_{-0.3}^{+0.3}$ \\
        $\Phi_2\,(^\circ)$        &$-$& $-$ & $1_{-120}^{+120}$\\
        \hline
    \end{tabular}
\end{table}

\subsection{Asymmetric Gaussian component} \label{sec: asymmetry}

\begin{figure}
    \centering
    \includegraphics[width=\columnwidth,trim={0cm 0 0 0},clip]{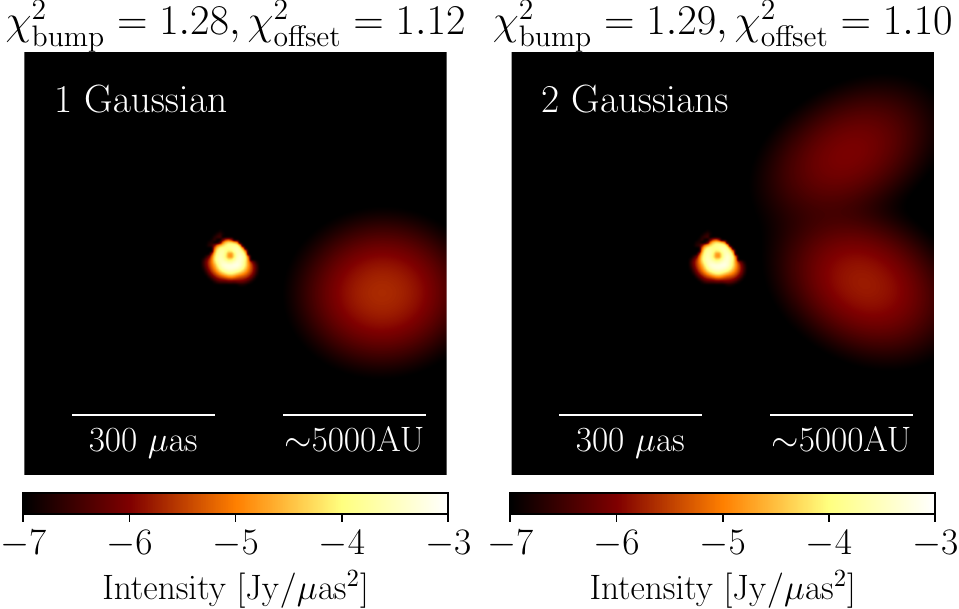}
    \caption{Images for the best fits with an asymmetric Gaussian component (left) and two Gaussian components (right) in log scale.}
    \label{fig: fiducial_toy_models}.
\end{figure}

To explore the possibility of asymmetry of the jet base emission we extend the model by two additional parameters: the ratio of $\textrm{fwhm}_\textrm{maj}$ and $\textrm{fwhm}_\textrm{min}$ ($\mathcal{R}$) and the rotation angle of the asymmetric Gaussian component $\Phi$, see Appendix~\ref{appendix: 6dimgauss} for additional details. We again use uniform priors ($\mathcal{R} \in [0, 1]$, and $\Phi \in [-180^\circ, 180^\circ]$). Neither the flux density nor the position change significantly; however, the model prefers to be symmetric, and hence the rotation angle is poorly constrained. The $\chi^2_\nu$ values are $1.28$ for the bump and $1.10$ for the offset, i.e., there is no significant improvement to the fit. The best-fit values are provided in \autoref{tab:params}, we show the best-fit model in the left panel of Figure~\ref{fig: fiducial_toy_models}.

\begin{figure*}
    \centering
    \includegraphics[width=\linewidth]{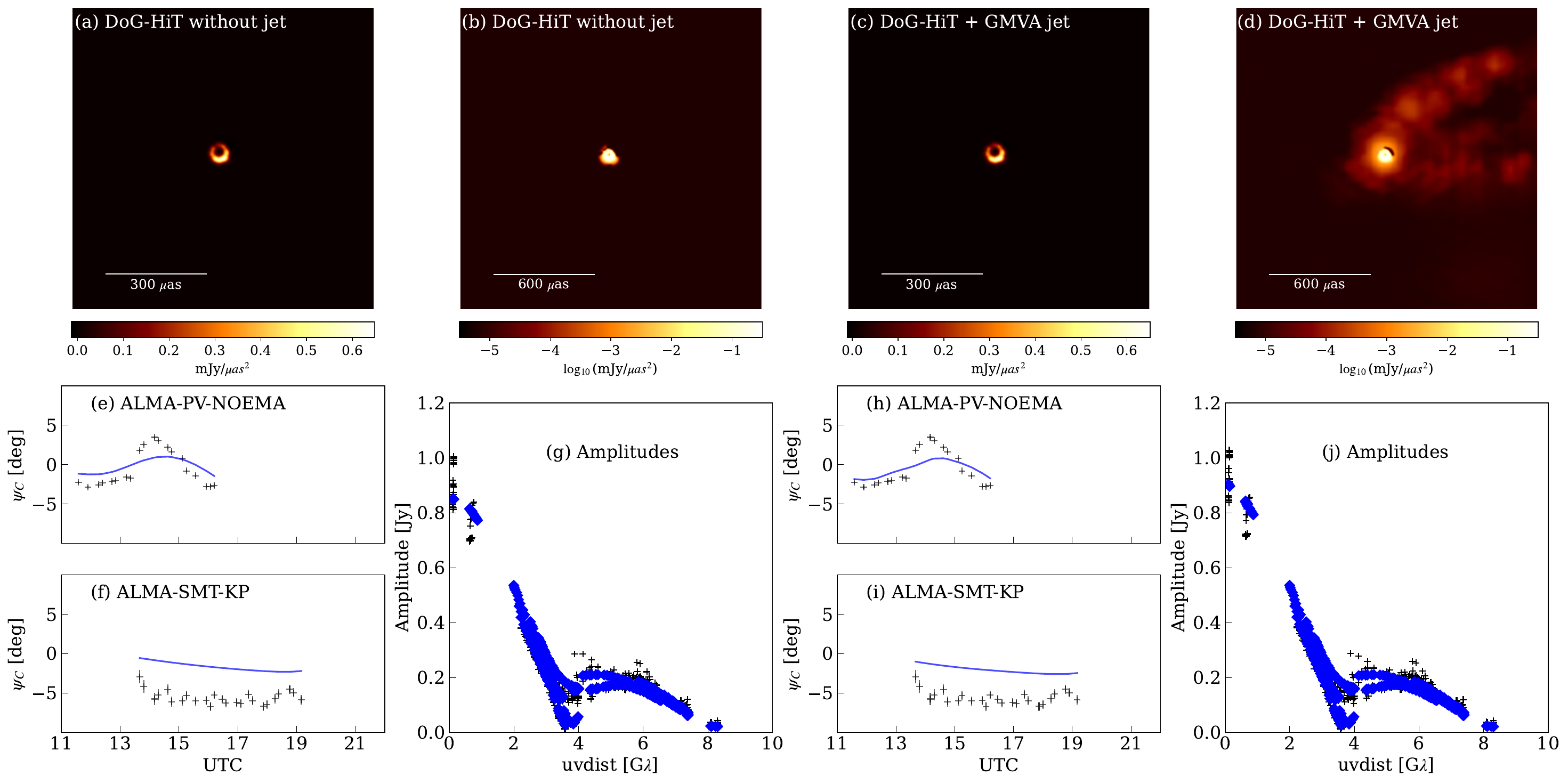}
    \caption{Ring-only \doghit image(first panel, top row: linear scale, second-panel first row: logarithmic scale) compared to the \doghit ring image stacked with the jet observed with the GMVA (third and fourth panel, top row). In the bottom row, we show the fit to the bump and offset in closure phases and the fit to the amplitudes for \doghit (left panels) and for \doghit with the GMVA jet added (right panels).}
    \label{fig: gmva_jet}
\end{figure*}
\subsection{Two Gaussians} \label{sec: 2gaussian}

For completeness, we explore the possibility of a secondary Gaussian component in the fit. However, given the complexity of the model and limitations of the data, we constrain the second component in the positive $y$ direction, that is, we allow the $y$ position to vary only within $[0, 500 \mu\textrm{as}]$. In this way, we test the presence of the second component in the northern region, but allow all parameters to vary ($12$ parameters). The flux density of the first southern Gaussian is $F_1=50_{-20}^{+30}\,\rm{mJy}$. The second Gaussian has a flux density of $F_2=20_{-10}^{+20}\,\rm{mJy}$. 
The first component is smaller in size compared to the previous modeling, but the positions do not vary significantly. The second, northern, component is relatively fainter. Since, the rotation angle for this component is poorly constrained, it prefers to be symmetric as well. The $\chi^2_\nu$ values are $1.29$ for the bump and $1.10$ for the offset. Thus, there is no statistical evidence for the inclusion of a secondary Gaussian component.

The Gaussian fit parameters are provided in Table~\ref{tab:params}, and the corner plot is provided in Appendix~\ref{appendix: corner plots}. The third image in Figure~\ref{fig: fiducial_toy_models} (right panel) represents the best-fit model consisting of two Gaussian components.

\section{Alternative fitting scenarios} \label{sec: alternatives}

Due to the limited number of data points on the two chosen closure triangles, multiple plausible emission features may exist. In the previous section, we demonstrated that adding a simple Gaussian component to the compact emission provides a good fit to the data; the emission is most likely located southwest of the ring emission. 

In this section, we discuss alternative models that may fit the data and discuss their validity. This list is not complete and focuses on some physically-motivated emission structures instead. Hence, we cannot rule out that alternative emission scenarios exist.

\subsection{GMVA jet structure}

At 86\,GHz, \cite{Lu2023} imaged a ring-like feature together with the innermost jet in \m87 with the GMVA complemented by the ALMA and the Greenland Telescope (GLT). Here, we discuss whether this jet structure would be able to fit the respective triangles. To this end, we need to extract the jet component of the GMVA image from the ring feature observed at 86\,GHz, and stack this jet feature to the ring feature observed at 230\,GHz. 

Specifically, \citet{Kim2024} presented a reanalysis of this data set, among others, with \doghit. In \doghit, the image structure is represented in multiple spatial scales by wavelets $\Psi_i$:
\begin{align}
    I = \sum_i \Psi_i * \omega_i. \label{eq: doghit}
\end{align}
The smallest-scale wavelets describe the ring (the small-scale feature), and the largest spatial scales describe the jet emission. In this way, the image is naturally decomposed into the compact ring structure and the jet base emission across different spatial scales. We use this natural decomposition to study whether the jet observed by the GMVA provides a good fit: we filter out the `ring-components' in the $86~\mathrm{GHz}$ image by setting all small-scale wavelets to zero, then we add the $230~\mathrm{GHz}$ EHT ring image and the jet emission at 86\,GHz seen by the GMVA. Motivated by the findings in Section~\ref{sec: impact_model}, the GMVA jet image is rescaled to add a flux density between $0.0\,\mathrm{Jy}$ and $0.2\,\mathrm{Jy}$ to match the 230\,GHz ring image. We grid-search for the best flux density in this interval and find the best $\chi^2$ at an additional flux density of $0.06\,\mathrm{Jy}$, consistent with the findings in Section~\ref{sec: fiducial}. 
The results are shown in Figure~\ref{fig: gmva_jet}. We note that we can absorb the large-scale jet component completely in the amplitude gains, providing a good fit to the self-calibrated amplitudes in both cases. The fit to the offset and the bump is improved by including the jet component. However, our model of the 86\,GHz jet is not able to provide a good fit, performing worse than the simpler fiducial model presented in Section~\ref{sec: fiducial}. This is likely a limitation of our attempt to scale the 86\,GHz image, rather than an indication that the 86\,GHz and 230\,GHz emission structures are inherently different.
\subsection{Imaging with weakened assumption on compactness}

Next, we check whether we can find an alternative way of fitting the data, including the offset and the bump utilizing the \doghit imaging algorithm. \doghit represents the recovered image by a dictionary of wavelet functions $\Psi_i$ and images at spatial sub-bands as described by Eq.~\ref{eq: doghit}. \doghit utilizes a sparsity-constraining approach, effectively defining multiresolution support, i.e., a set of statistically-significant wavelet coefficients to describe the image. As demonstrated in \citet{Mueller2023b}, the multiresolution support encodes two constraints: the location of the emission and the spatial scales needed to explain the emission at these locations. The latter relates to the \uv coverage, i.e., which spatial scales are measured. 

This way, \doghit allows for three alternative approaches to fit the bump and the offset. In the first approach, we investigate whether large spatial scales can account for the observed deviations without a constraint on the emission location or its spatial scale. In the second approach, we focus on localized emission in the ring. In the third approach, we constrain the location and spatial scales. For each approach, we take the ring-only image from \doghit and try to improve the fit quality to the closure phases only, using a gradient descent approach with a small step size.

\subsubsection{Fit without spatial scale and localization constraints}

\begin{figure}
    \centering
    \includegraphics[width=\linewidth]{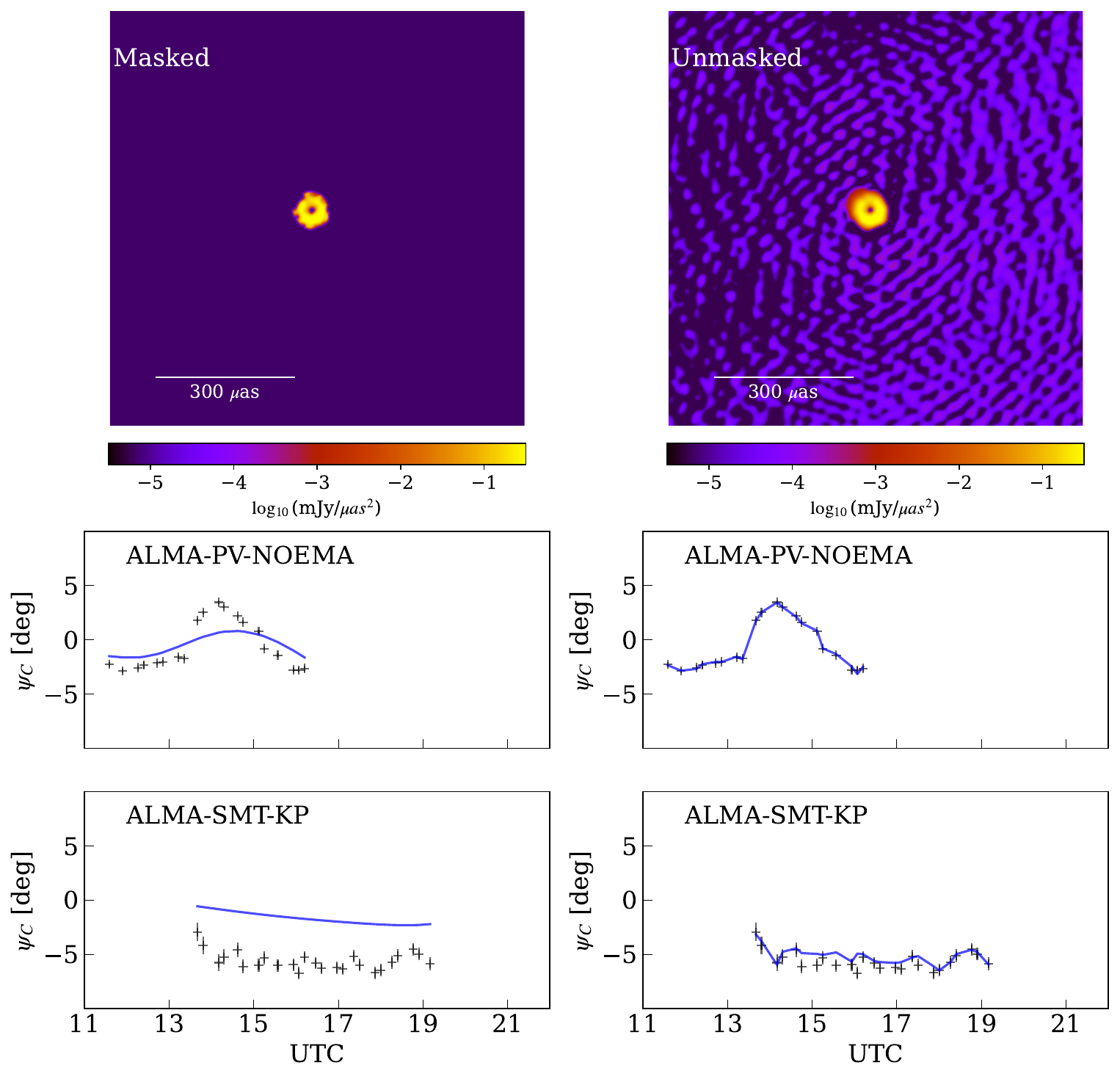}
    \caption{Reconstruction of the ring and potential jet base emission, when restricted to spatial scales determined by \doghit in the compact emission region (left panels, called masked), and completely unmasked (right panels, called unmasked).}
    \label{fig: masked_unmasked_zoom}
\end{figure}

First, we do this fit without any mask on the spatial scales or the location of the emission, allowing every pixel to vary (labeled unmasked Figure~\ref{fig: masked_unmasked_zoom}, right column). The unmasked image fits the bump and also shows a much-improved fit to the offset. Moreover, improving the fits in the direction of the closure phases does not violate the match to the amplitudes. The recovered ring image shows no significant variation. However, comparing the unmasked image with the original image, which was masked in spatial scale and localization (Figure~\ref{fig: masked_unmasked_zoom}, left column), on a logarithmic scale, we observe that the improved precision has been achieved by fitting a `waffle-like pattern' associated with the PV-NOEMA and SMT-KP baselines, which is nonphysical. 

\subsubsection{Fit with localization constraint}

\begin{figure}
    \centering
    \includegraphics[width=\linewidth]{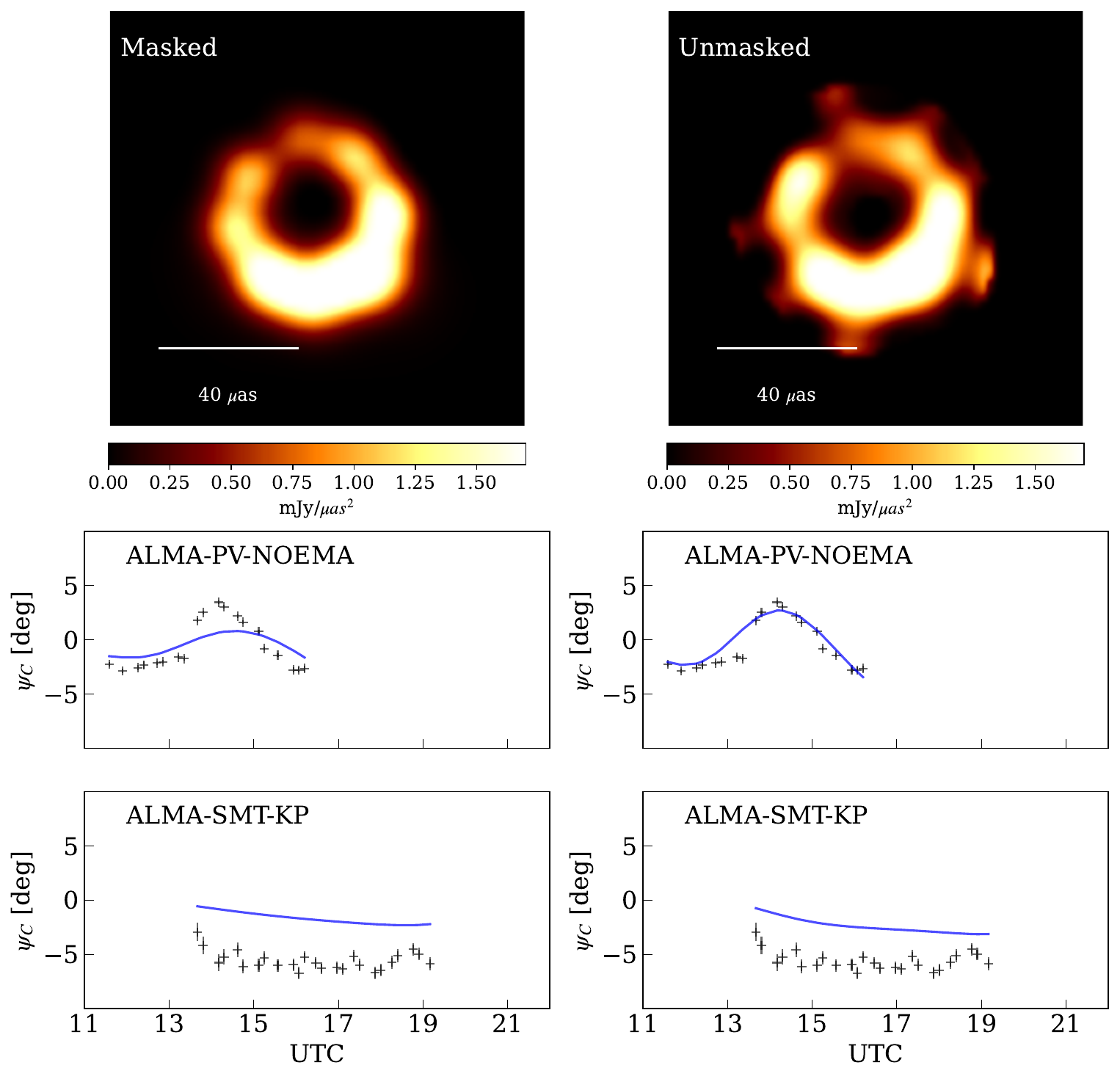}
    \caption{Reconstruction of the ring and potential jet base emission when restricted to spatial scales determined by \doghit in the compact emission region (left panels, called masked), and when restricted to the ring emission, but flexible in spatial scale (right panels, called unmasked).}
    \label{fig: masked_unmasked_single}
\end{figure}

Here, we add a constraint on the emission location by performing the same gradient descent fit to the closure phases but only allowing the pixels on the ring to vary, defined by the $1\%$ intensity contour in the \doghit imagesy. The resulting reconstructions and model closure phases are shown in Figure~\ref{fig: masked_unmasked_single}.

In this case, we can fit the bump reasonably well but have issues fitting the offset. A closer inspection of the structures that fit the bump in the top right panel of Figure~\ref{fig: masked_unmasked_single} indicates that this fitting is achieved by asymmetries in the ring itself through a number of components. Each component exhibits structure significantly smaller than the resolution limit of roughly $20\,\mathrm{\mu as}$.

\subsubsection{Fit with localization and spatial constraints}

Finally, we constrain the emission both by its localization and its spatial scale, as imprinted in the wavelet approach of \doghit. The respective fitting quality is shown in the left columns of Figure~\ref{fig: masked_unmasked_zoom} and Figure~\ref{fig: masked_unmasked_single}. Even if we fit for the ALMA-PV-NOEMA and ALMA-KP-SMT closure phases alone with a gradient-descent approach, restricted by the localization and spatial constraints imprinted by the wavelet dictionary, we remain limited in our potential to fit the bump or the offset.

This suggests that the model mismatch results from limiting the imaging to ring-scale emission only.
This further supports the view that the bump and the offset are signatures of jet base emission, rather than artifacts from the imaging procedure being imprecise in its representation of the ring.

\subsubsection{Data corruptions and systematic errors}

Both the bump and offset are formally significant relative to a $0\degree$ closure phase, assuming the thermal errorbars are reliable. However, it is likely that the thermal errorbars underestimate the true error, as they do not account for non-closing errors such as polarization leakage \citep[e.g.,][]{Broderick2020_closure}. Therefore, imaging methods add a fractional systematic Gaussian uncertainty to the errors of $\sim 2\%$ of the visibility amplitudes \citep{M87p3}. In \autoref{app:leakage} we demonstrate that leakage cannot account for the bump or the offset. Nevertheless, we cannot rule out the possibility that other unknown sources of systematic error may corrupt the data.

\subsection{Impact on previous EHT observations}

The information about jet base emission has been derived primarily from the KP-SMT and PV-NOEMA baselines in this manuscript. These baselines are unique to the 2021 EHT data, as this was the first year in which the Kitt Peak telescope and NOEMA participated in EHT observations \citep{M87_2021}. In 2017 and 2018, LMT formed a baseline with the SMT of $\sim1.3\,\mathrm{G}\lambda$ (compared to $0.8\,\mathrm{G}\lambda$ for PV-NOEMA in 2021), making it the shortest baseline between two stations that are not co-located (ALMA/APEX and SMA/JCMT) during these years. However, LMT did not participate in the 2021 observations. Here, we examine whether the jet base component derived in this manuscript -- based on the enhanced coverage of the EHT in 2021 -- would have a significant impact on data taken in 2017 and 2018. In other words, does our knowledge of an extended component influence the black hole shadow reconstructions presented in \citet{M87p1, M87_2018p1, M87_2021}?

\begin{figure}
    \centering
    \includegraphics[width=\linewidth]{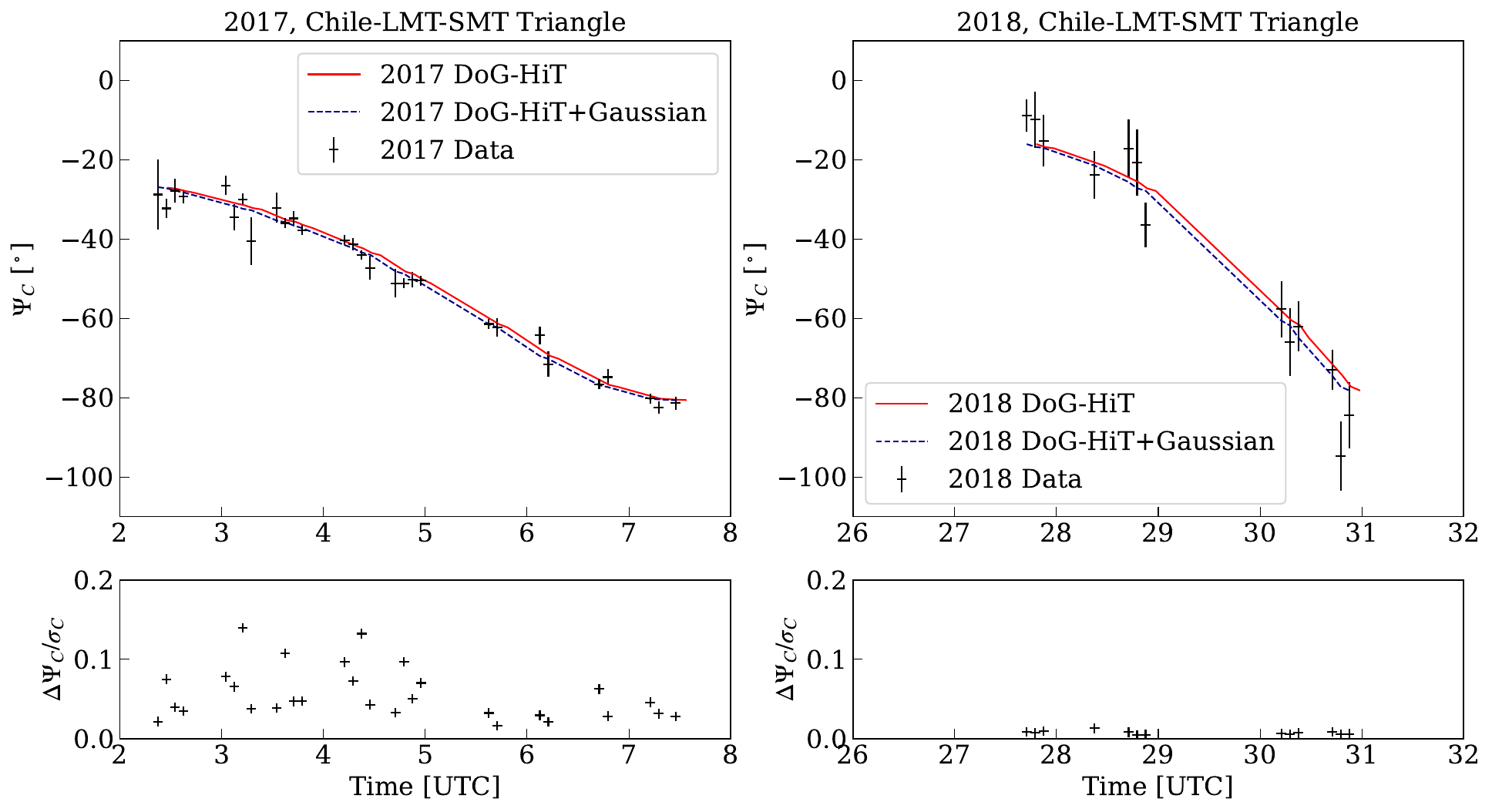}
    \caption{Closure triangle formed by Chilean stations (ALMA/APEX), LMT, and SMT in 2017 (left panel) and 2018 (right panel). The LMT-SMT baseline is the shortest baseline between non-co-located stations in EHT observations of 2017 and 2018. Black markers: observed closure phases. Red line: fit to the \doghit reconstruction in 2017 (left panel) and 2018 (right panel). Dark blue line: fit to the \doghit reconstruction (ring) with an additional Gaussian model component determined from 2021 data as it would appear in 2017 (left panel) and 2018 (right panel). The line has been offset by a time offset of $0.1\,\mathrm{hours}$ to make it visible. In the bottom panels, we show the significance of the difference between the \doghit model and the \doghit model with additional Gaussian component, measured by the difference in $\Psi_C$ divided by the closure phase error. In both years, the effect of the additional Gaussian component on the Chile-LMT-SMT closure phases is an order of magnitude smaller than the respective noise.}
    \label{fig: LMTtriangle}
\end{figure}

In \autoref{fig: LMTtriangle}, we show the closure phases observed in 2017 (left panel) and 2018 (right panel) with the LMT-SMT baseline (black data points), and their respective fit to the \doghit model of that year presented in \citet{M87_2021} (red line). The closure phases undergo a swing of almost $80^\circ$ over the full duration of the observing track. This swing is completely explained by the ring models recovered by \doghit. There is no additional feature visible that would motivate the fitting of an additional component in these years from the LMT-SMT baseline. In addition, we show with a blue dotted line the closure phases of the \doghit model with the additional Gaussian component obtained from our analysis in 2021. In the bottom panels, we show the difference between the predicted closure phases with and without the additional Gaussian component compared to the noise level. The difference in closure phase introduced by our model on the LMT-SMT baseline is negligible. We conclude that, due to the lack of short baselines, the data obtained in 2017 and 2018 by the EHT are fully fitted by ring-only emission. Additionally, the Gaussian component identified in this work is not detectable on any baseline in the EHT longer than the PV-NOEMA baseline, and thus was not detectable/had no measurable impact on observations conducted in 2017 and 2018. Moreover, the component fitted here is relatively faint, only $60\,\mathrm{mJy}$. Any significantly brighter component would introduce effects in ALMA-PV-NOEMA and ALMA-KP-SMT triangles larger than a few degrees, setting a robust upper limit for the amount of resolved jet emission in EHT observations to date.

\begin{figure}
    \centering
    \includegraphics[width=\columnwidth]{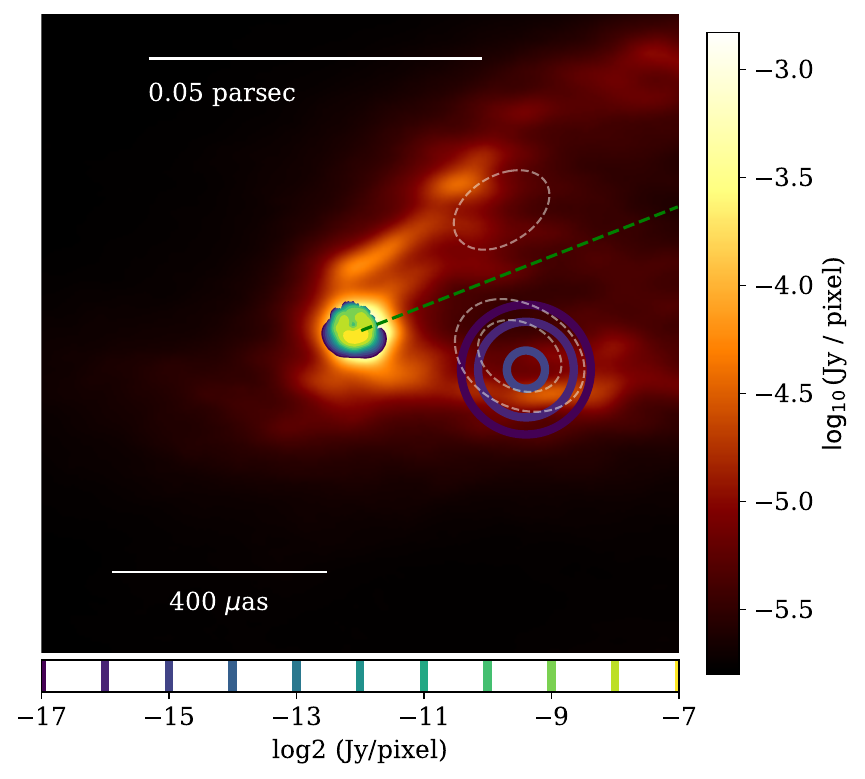}
    \caption{Fiducial ring structure of the \m87 black hole shadow (green contour) with a single Gaussian component (solid blue contours) corresponding to the best-fit parameters found in Section~\ref{sec: fiducial} (Table~\ref{tab:params}, first set) and double Gaussian components (dashed contours) corresponding to the best-fit parameters found in Section~\ref{sec: 2gaussian} (Table~\ref{tab:params}, second set), overplotted on the jet in \m87 observed at $86\, \mathrm{GHz}$ (heatmap) with the GMVA \citep{Lu2023, Kim2024}. The green dotted line shows the jet orientation.}
    \label{fig: eht_gmva}
\end{figure}
\section{Discussion}
\citet{Broderick2022} found a component roughly $50\,\mu\mathrm{as}$ to the south and $50\,\mu\mathrm{as}$ to the West of the black hole shadow by modeling a narrow ring component to describe the black hole ring feature in the 2017 EHT observations \citep{M87p1}. This result is consistent with similar findings reported in \citet{Arras2022, Carilli2022}, although a robust detection could not be claimed by either study due to the lack of short and intermediate baselines in the 2017 EHT observations.

\citet{Broderick2022} demonstrated that the feature aligns well with the southern arm of the edge brightened jet observed at 86\,GHz \citep{Kim2018} when taking into account the expected core shift between 230 and 86\,GHz. Recently \citet{Lu2023} observed the innermost jet in \m87 at 86\,GHz with the GMVA expanded by ALMA. They found a ring-like feature with an edge-brightened jet base. In Figure~\ref{fig: eht_gmva} we present the EHT results (green-colored contours) on top of the GMVA image (in virdis colormap). To this end, we use the intensity map reported by \citealt{Kim2024} (a reanalysis of results reported in \citealt{Lu2023}) due to its improved resolution. The images are stacked by aligning the ring features observed at 230 and 86\,GHz. By visual inspection, the jet base Gaussian component studied here aligns to first order with the southern arm of the edge-brightened jet observed with the GMVA. The second Gaussian component in the two-Gaussian model would align with the northern arm, but most likely represents a degeneracy in the posterior.

The jet collimation profile at sub-milliarcsecond scales is parabolic. \citet{Kim2018} found a jet width profile $W \propto z^{0.5}$ (where $z$ is the projected distance from the core), also in agreement with observations at milliarcsecond scales \citep{Hada2013, Hada2016, Hada2017}. This parabolic shape is consistent with a jet driven either by black hole spin via the Blandford-Znajek mechanism \citep{Blandford1977, Nakamura2018} or by the differential rotation of the accretion flow via the Blandford-Payne mechanism \citep{blandford-payne}. The preference for a southwestern component in our model over a northwestern component is consistent with a rapidly rotating jet structure near the black hole \citep{Broderick2022}, however the scope for interpretation remains limited due to the limited amount of information. \citet{Lu2023} reported a deviation from the parabolic shape near the core ($\leq0.2\,\mathrm{mas}$).  This profile flattening may be a powerful indicator of the jet launching mechanism, potentially requiring an additional emission component, for example by non-thermal electrons in gravitationally unbound, non-relativistic winds \citep{Nakamura2018, Park2019, Lu2023}. While this is an anticipated observable for the EHT and its successors to connect the jet dynamics to the black hole and study jet launching mechanisms \citep{Midtermgoals}, we note that the jet base component found in this study is too far out to make reliable conclusions about the physics of the jet within the first ten Schwarzschild radii.

\section{Conclusions}

We used the improved intermediate baseline coverage of the 2021 EHT campaign to search for resolved jet base emission around \m87. We found evidence for a faint offset component on angular scales accessible to the PV-NOEMA and KP-SMT baselines. Fitting a simple geometric model -- a single offset Gaussian -- yields a flux density of $\sim 60~\mathrm{mJy}$ at a location $\approx (-60,-320)~\mathrm{\mu as}$ relative to the ring center. The Gaussian component has $\mathrm{fwhm} \approx 180\,\mu\rm{as}$.

Introducing the Gaussian component into the source model reduces the closure phase residuals on the ALMA-PV-NOEMA and ALMA-KP-SMT triangles. For the ALMA-PV-NOEMA triangle the $\chi^2_\nu$ falls from $2.73$ to $1.29$, and for the ALMA-KP-SMT triangle from $2.52$ to $1.12$; the $\chi^2_\nu$ for all triangles improves from $1.50$ to $1.39$; and it is improved on every single triangle, albeit the improvement is insignificant for most. This demonstrates that a compact, faint asymmetry at intermediate scales is an economical explanation for the observed closure phase structure in the 2021 data.

The recovered component’s position and scale are consistent with the southern/southwestern arm of the edge-brightened jet seen in the 86\,GHz GMVA image. Stacking the component over the GMVA jet indicates a preference for an additional flux density of $\sim60 \mathrm{mJy}$ to match the EHT closure phases. However, the data do not statistically favor more complex models (an elliptical Gaussian component or a two-component component model) over the single Gaussian model.

We emphasize two important caveats. First, while our modeling improves the data-minus-model residuals, the limited number of intermediate baselines and the single-baseline sensitivity at the relevant spatial frequencies restrict a robust morphological reconstruction: other mathematically viable structures, including imaging artifacts consistent with the measured \uv sampling, cannot be fully excluded. Second, single-baseline or small-triangle measurements are susceptible to non-closing systematics. While we rule out polarization leakage as the origin of the signal and have performed imaging and synthetic data checks, unknown systematic errors cannot be completely ruled out. For these reasons, we conservatively treat the recovered Gaussian flux density as an upper limit on resolved jet base emission at spatial scales probed by the PV–NOEMA and KP–SMT baselines.

Finally, the presence of this faint component does not conflict with earlier EHT reconstructions from 2017, 2018, and 2021. The arrays in earlier years lacked the intermediate baselines required to detect emission at the spatial scales we probe here, so ring-only models adequately describe the earlier epochs. Excellent fitting quality, synthetic data tests, and the negligible effect of the Gaussian model on baselines longer than PV-NOEMA also indicate the robustness of the ring-only images obtained in 2021. The detection (or upper limit) reported here therefore complements earlier results and constrains where the EHT missing flux cannot reside: resolved, brighter emission $\sim 100 - 300\,\mu\mathrm{as}$ southeast and northeast relative to the ring is firmly ruled out by these closure triangles.

In summary, the 2021 EHT data provide the first constraints at 230\,GHz on faint, asymmetric emission at intermediate scales near \m87, locating a plausible component southwest of the ring with a flux density $< 0.1\,\mathrm{Jy}$. Although this result is robust under the assumptions and tests performed, definitive confirmation and more precise constraints will require future EHT observations with higher sensitivity and improved intermediate-baseline coverage via additional stations and expanded frequency range. 

\bibliographystyle{aa}
\bibliography{references}{}

\begin{appendix}
\section{Corner plots}\label{appendix: corner plots}

\begin{figure}
    \includegraphics[width=\columnwidth]{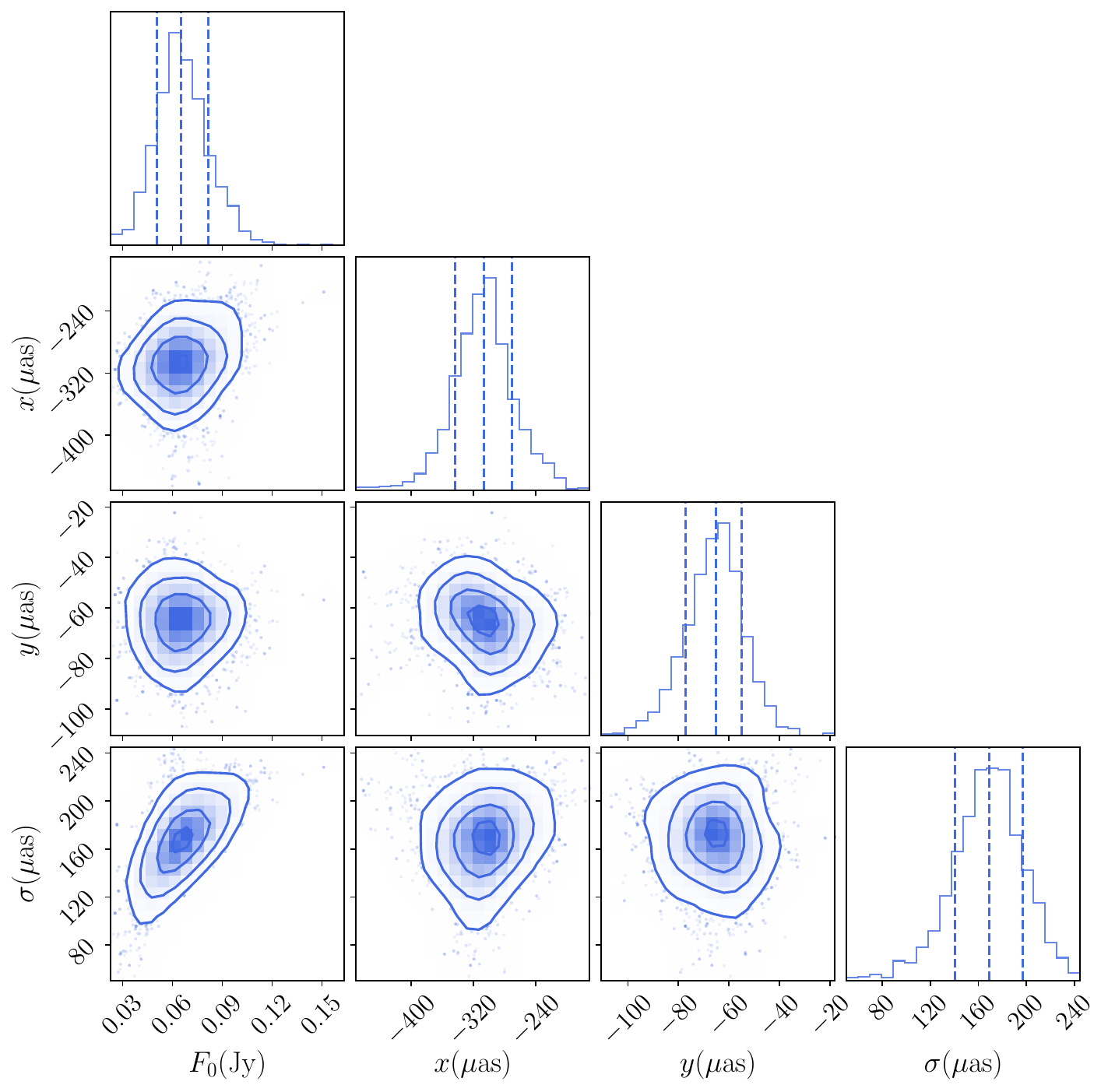}
    \caption{Corner plot showcasing our estimates for the location ($x,y$ in units of $\mu\,\mathrm{as}$), flux ($F_0$ in units of $\mathrm{Jy}$), and width ($\sigma$ in units of $\mu\,\mathrm{as}$) of the Gaussian toy model.}
    \label{fig: cornerplot}
\end{figure}

\begin{figure}
    \includegraphics[width=\columnwidth]{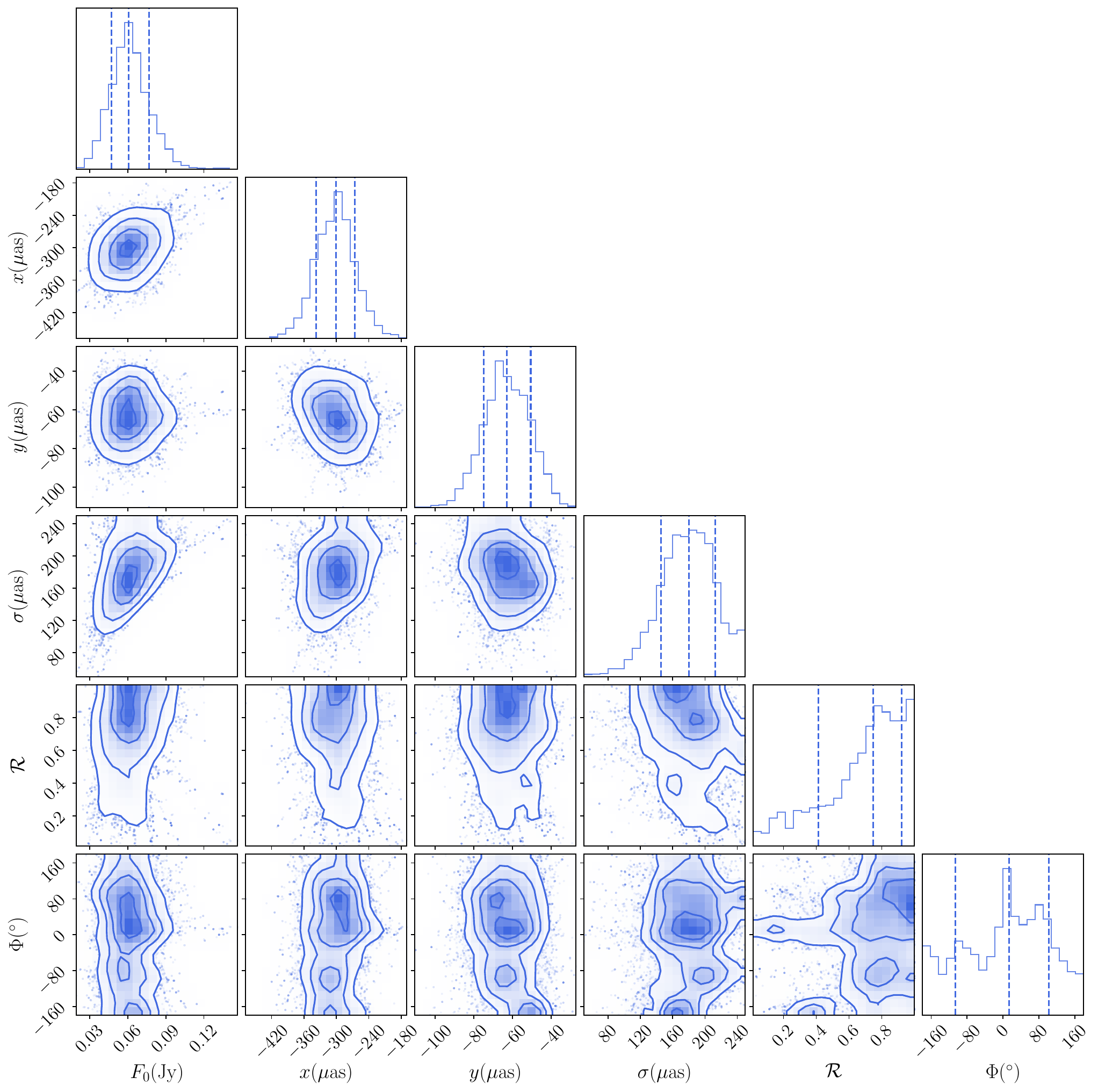}
    \caption{ Same as Figure~\ref{fig: cornerplot} but with two additional parameters: $\mathcal{R}$, the ratio of $\textrm{fwhm}_\textrm{maj}$ and $\textrm{fwhm}_\textrm{min}$, and $\Phi$ , the rotation angle, of the asymmetric Gaussian toy model. See Section~\ref{sec: asymmetry} for more details.}
    \label{fig: cornerplot2}
\end{figure}

\begin{figure}
    \includegraphics[width=\columnwidth]{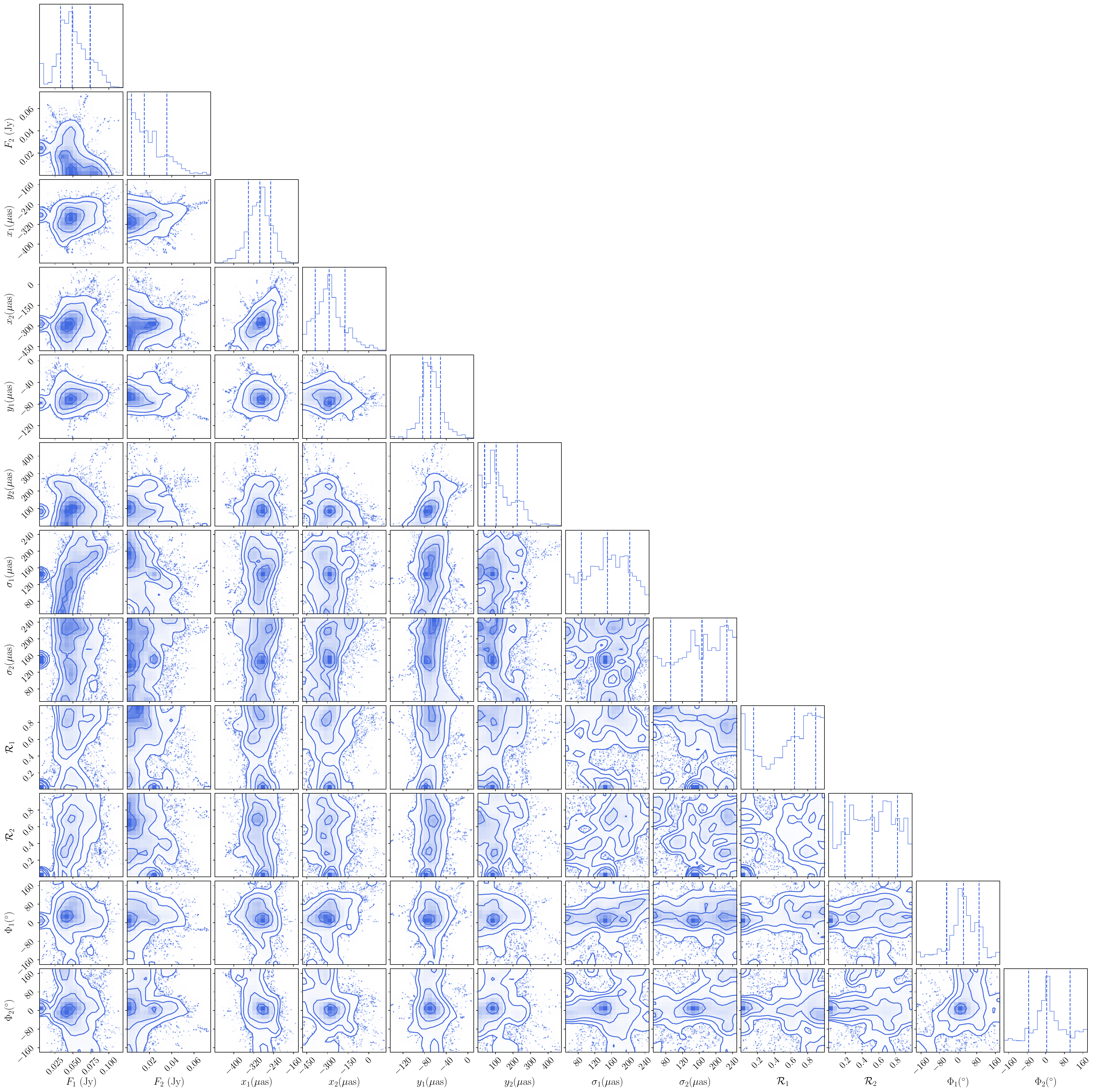}
    \caption{Same as Figure~\ref{fig: cornerplot} but for two asymmetric Gaussian toy models. See Section~\ref{sec: 2gaussian} for more details.}
    \label{fig: cornerplot3}
\end{figure}

In the following, we show the MCMC corner plots of the three model fits that we performed. Figure~\ref{fig: cornerplot} displays the corner plot of the four-parameter Gaussian model, Figure~\ref{fig: cornerplot2} displays the corner plot of the six-parameter Gaussian model, and Figure~\ref{fig: cornerplot3} displays the fit with a twelve-parameter double Gaussian model.


\section{Effect on visibility amplitude}\label{appendix: visbilities}
\begin{figure}
    \includegraphics[width=\columnwidth]{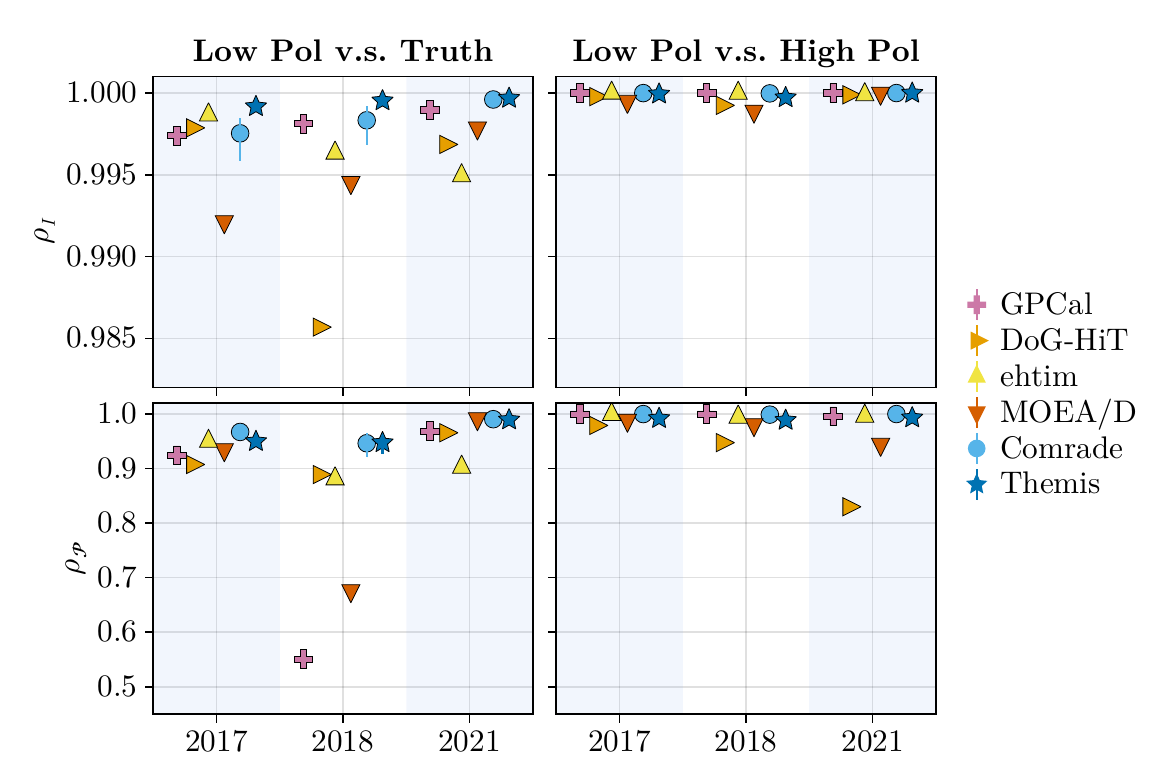}
    \caption{Figure~D3 of \citet{M87_2021}. Cross-correlation between the recovered ring model and synthetic ground truth model for seven different imaging techniques in 2017, 2018, and 2021. The synthetic data included a large-scale jet component. Upper panels show the cross-correlation in total intensity, lower panels in polarization. Despite the synthetic data containing a large scale jet component, the black hole shadow has been recovered for all methods and all years with exceptional accuracy, achieving cross-correlations $>0.995$ in total intensity in 2021.}
    \label{fig: jet_test}
\end{figure}

\begin{figure}
    \includegraphics[width=\columnwidth]{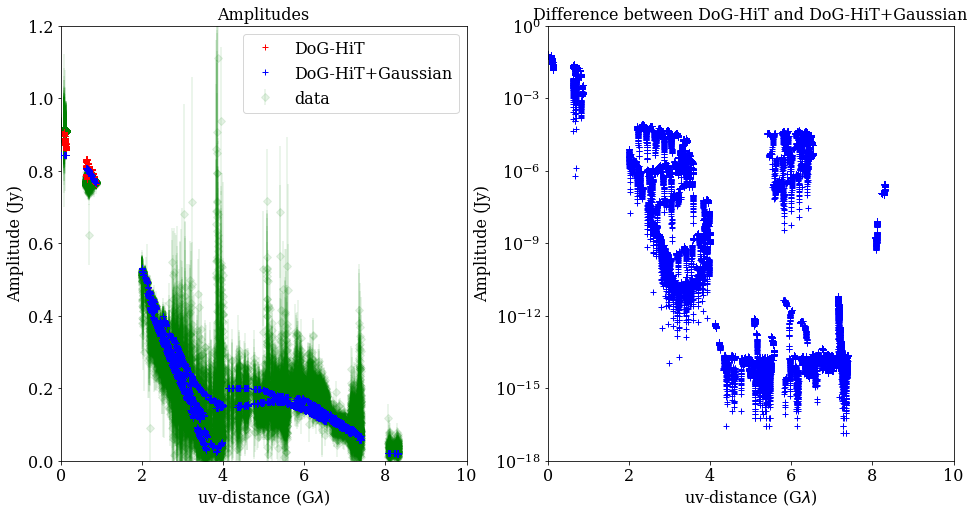}
    \caption{Left panel: comparison of the predicted amplitudes for the \doghit ring-only image (red markers), the \doghit image with additional Gaussian component (blue markers), and the self-calibrated visibility amplitudes (green markers) as a function of projected baseline length. Right panel: the difference between the predicted visibilities for the \doghit model and \doghit with the additional Gaussian component as a function of projected baseline length. The plot demonstrates that the additional Gaussian component has no sizeable effect on baselines longer than PV-NOEMA.}
    \label{fig: amplitudes}
\end{figure}

In Figure~\ref{fig: jet_test} we show the evaluation of the synthetic data test performed with an extended jet component in \citet{M87_2021}. In Figure~\ref{fig: amplitudes} we show the amplitudes of the ring model, the fiducial model, and the difference between the two models as a function of the projected baseline length.

\begin{figure}[!ht]
    \includegraphics[width=\columnwidth]{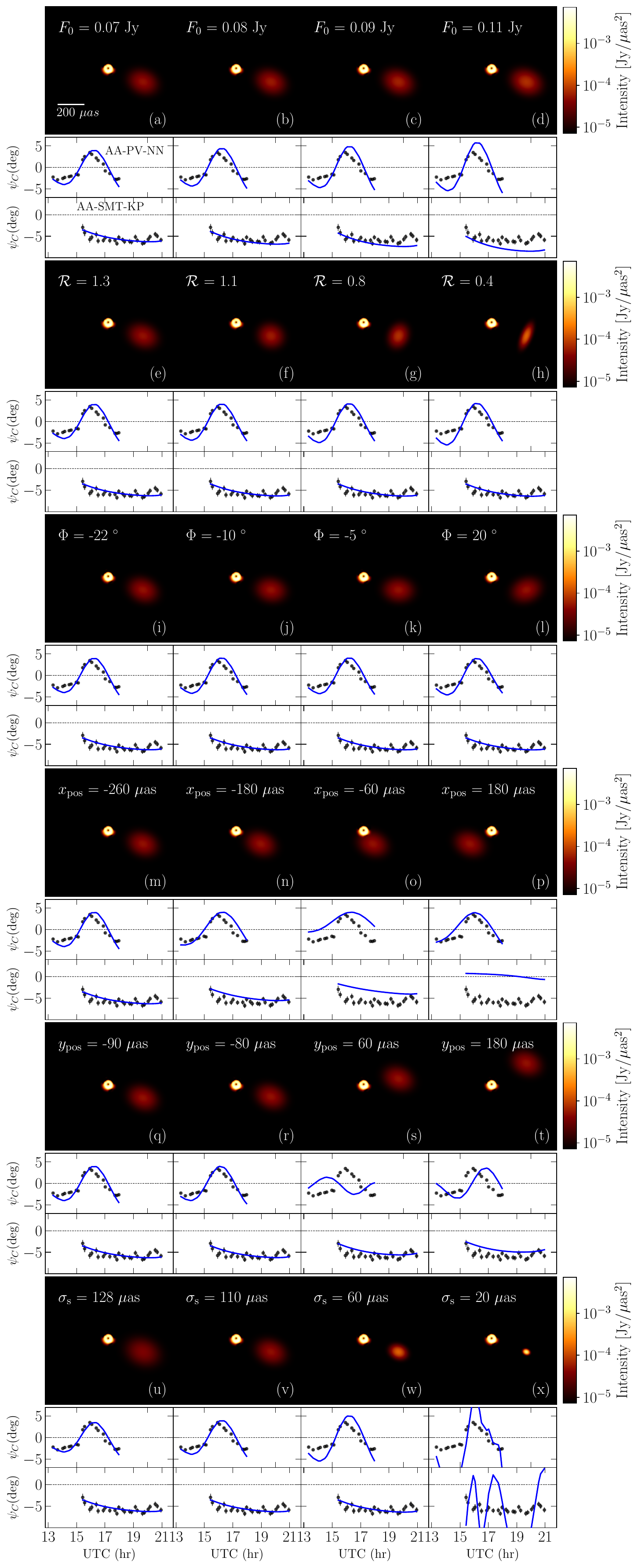}
    \caption{Same as Figure~\ref{fig: toy_gauss}, but with asymmetry and position angle as additional parameters.}
    \label{fig: toy_gauss_full}
\end{figure}
\section{Parameter exploration for 6 parameter Gaussian model}\label{appendix: 6dimgauss}

We demonstrate the effect of adding two additional degrees of freedom to the model with an asymmetric Gaussian component that can be rotated around its axis. 
In Figure~\ref{fig: toy_gauss_full} we show that these parameters can be constrained in the extreme case of a very asymmetric Gaussian component. However, these parameters are poorly constrained.
\section{Impact of non-closing errors}\label{app:leakage}
\begin{figure}[t]
    \includegraphics[width=\columnwidth]{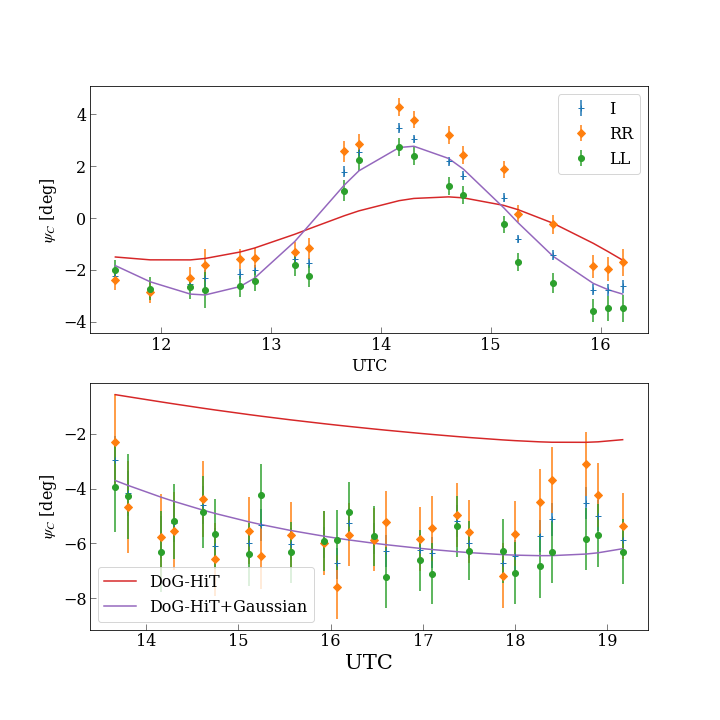}
\caption{Comparison of the $RR$ (orange) and $LL$ (green) closure phases on the ALMA-NOEMA-PV (top) and ALMA-KP-SMT triangles (bottom). The red line indicates the \doghit ring model, the magenta line indicates the \doghit model with an additional Gaussian component.}
\label{fig: leakage}
\end{figure}

Because the phase signature of the offset Gaussian component(s) is small, we cannot completely rule out systematic effects as an origin. Here, we assess the impact of one known non-closing error: polarization leakage.
Polarization leakage causes signal dispersion from one correlation of polarization handedness (e.g., $RR$) to another (e.g., $LL$), and is baseline-based in nature \citep[e.g.,][]{Mueller2024b}. One data-driven test to rule out leakage as the cause of the bump and offset is to compare the phase-residuals of the $RR-LL$ closure phases on the respective triangles (\autoref{fig: leakage}). 
Under the assumptions that there is no significant leakage and that circular polarization is $\xrightarrow{}0$, the difference between the parallel hands closure phases is zero.
While there is a small offset on the ALMA-PV-NOEMA triangle, neither the bump nor the offset can be explained by a difference between the parallel hands, and we thus conclude that the signal is not caused by polarization leakage. The difference in the ALMA-PV-NOEMA triangle may be caused by either intrinsic circular polarization of the source or by leakage, and we thus place a limit on this type of non-closing error of $\sim 1\mathrm{\degree}$.
\section{Acknowledgments}
\begin{acknowledgements}

We thank the anonymous reviewer for their insightful comments and suggestions.
Saurabh received financial support for this research from the International Max Planck Research School (IMPRS) for Astronomy and Astrophysics at the Universities of Bonn and Cologne. This work was supported by the M2FINDERS project funded by the European Research Council (ERC) under the European Union's Horizon 2020 Research and Innovation Programme (Grant Agreement No. 101018682). SDvF gratefully acknowledges the support of the Alexander von Humboldt Foundation through a Feodor Lynen Fellowship and thanks CITA for their hospitality and collaboration. 
The Event Horizon Telescope Collaboration thanks the following
organizations and programs: the Academia Sinica; the Research Council of Finland (project 362572); the Agencia Nacional de Investigaci\'{o}n 
y Desarrollo (ANID), Chile via NCN$19\_058$ (TITANs), Fondecyt 1221421 and BASAL FB210003; the Alexander
von Humboldt Stiftung (including the Feodor Lynen Fellowship); an Alfred P. Sloan Research Fellowship;
Allegro, the European ALMA Regional Centre node in the Netherlands, the NL astronomy
research network NOVA and the astronomy institutes of the University of Amsterdam, Leiden University, and Radboud University;
the ALMA North America Development Fund; the Astrophysics and High Energy Physics programme by MCIN (with funding from European Union NextGenerationEU, PRTR-C17I1); the Black Hole Initiative, which is funded by grants from the John Templeton Foundation (60477, 61497, 62286) and the Gordon and Betty Moore Foundation (Grant GBMF-8273) - although the opinions expressed in this work are those of the author and do not necessarily reflect the views of these Foundations; 
the Brinson Foundation; the Canada Research Chairs (CRC) program; Chandra DD7-18089X and TM6-17006X; the China Scholarship
Council; the China Postdoctoral Science Foundation fellowships (2020M671266, 2022M712084); ANID through Fondecyt Postdoctorado (project 3250762); Conicyt through Fondecyt Postdoctorado (project 3220195); Consejo Nacional de Humanidades, Ciencia y Tecnología (CONAHCYT, Mexico, projects U0004-246083, U0004-259839, F0003-272050, M0037-279006, F0003-281692, 104497, 275201, 263356, CBF2023-2024-1102, 257435); the Colfuturo Scholarship; the Consejo Superior de Investigaciones 
Cient\'{i}ficas (grant 2019AEP112);
the Delaney Family via the Delaney Family John A.
Wheeler Chair at Perimeter Institute; Dirección General de Asuntos del Personal Académico-Universidad Nacional Autónoma de México (DGAPA-UNAM, projects IN112820 and IN108324); the Dutch Research Council (NWO) for the VICI award (grant 639.043.513), the grant OCENW.KLEIN.113, and the Dutch Black Hole Consortium (with project No. NWA 1292.19.202) of the research programme the National Science Agenda; the Dutch National Supercomputers, Cartesius and Snellius  (NWO grant 2021.013); 
the EACOA Fellowship awarded by the East Asia Core
Observatories Association, which consists of the Academia Sinica Institute of Astronomy and Astrophysics, the National Astronomical Observatory of Japan, Center for Astronomical Mega-Science,
Chinese Academy of Sciences, and the Korea Astronomy and Space Science Institute; 
the European Research Council (ERC) Synergy Grant ``BlackHoleCam: Imaging the Event Horizon of Black Holes'' (grant 610058) and Synergy Grant ``BlackHolistic:  Colour Movies of Black Holes:
Understanding Black Hole Astrophysics from the Event Horizon to Galactic Scales'' (grant 10107164); 
the European Union Horizon 2020
research and innovation programme under grant agreements
RadioNet (No. 730562), 
M2FINDERS (No. 101018682); the European Research Council for advanced grant ``JETSET: Launching, propagation and 
emission of relativistic jets from binary mergers and across mass scales'' (grant No. 884631); the European Horizon Europe staff exchange (SE) programme HORIZON-MSCA-2021-SE-01 grant NewFunFiCO (No. 10108625); the Horizon ERC Grants 2021 programme under grant agreement No. 101040021; the FAPESP (Funda\c{c}\~ao de Amparo \'a Pesquisa do Estado de S\~ao Paulo) under grant 2021/01183-8; the Fondes de Recherche Nature et Technologies (FRQNT); the Fondo CAS-ANID folio CAS220010; the Generalitat Valenciana (grants APOSTD/2018/177 and  ASFAE/2022/018) and
GenT Program (project CIDEGENT/2018/021); the Gordon and Betty Moore Foundation (GBMF-3561, GBMF-5278, GBMF-10423);   
the Institute for Advanced Study; the ICSC – Centro Nazionale di Ricerca in High Performance Computing, Big Data and Quantum Computing, funded by European Union – NextGenerationEU; the Istituto Nazionale di Fisica
Nucleare (INFN) sezione di Napoli, iniziative specifiche
TEONGRAV; 
the International Max Planck Research
School for Astronomy and Astrophysics at the
Universities of Bonn and Cologne; the Italian Ministry of University and Research (MUR)– Project CUP F53D23001260001, funded by the European Union – NextGenerationEU; 
Deutsche Forschungsgemeinschaft (DFG) research grant ``Jet physics on horizon scales and beyond'' (grant No. 443220636) and DFG research grant 443220636;
Joint Columbia/Flatiron Postdoctoral Fellowship (research at the Flatiron Institute is supported by the Simons Foundation); 
the Japan Ministry of Education, Culture, Sports, Science and Technology (MEXT; grant JPMXP1020200109); 
the Japan Society for the Promotion of Science (JSPS) Grant-in-Aid for JSPS
Research Fellowship (JP17J08829); the Joint Institute for Computational Fundamental Science, Japan; the Key Research
Program of Frontier Sciences, Chinese Academy of
Sciences (CAS, grants QYZDJ-SSW-SLH057, QYZDJSSW-SYS008, ZDBS-LY-SLH011); 
the Leverhulme Trust Early Career Research
Fellowship; the Max-Planck-Gesellschaft (MPG);
the Max Planck Partner Group of the MPG and the
CAS; the MEXT/JSPS KAKENHI (grants 18KK0090, JP21H01137,
JP18H03721, JP18K13594, 18K03709, JP19K14761, 18H01245, 25120007, 19H01943, 21H01137, 21H04488, 22H00157, 23K03453); the MICINN Research Projects PID2019-108995GB-C22, PID2022-140888NB-C22; the MIT International Science
and Technology Initiatives (MISTI) Funds; 
the Ministry of Science and Technology (MOST) of Taiwan (103-2119-M-001-010-MY2, 105-2112-M-001-025-MY3, 105-2119-M-001-042, 106-2112-M-001-011, 106-2119-M-001-013, 106-2119-M-001-027, 106-2923-M-001-005, 107-2119-M-001-017, 107-2119-M-001-020, 107-2119-M-001-041, 107-2119-M-110-005, 107-2923-M-001-009, 108-2112-M-001-048, 108-2112-M-001-051, 108-2923-M-001-002, 109-2112-M-001-025, 109-2124-M-001-005, 109-2923-M-001-001, 
110-2112-M-001-033, 110-2124-M-001-007 and 110-2923-M-001-001); the National Science and Technology Council (NSTC) of Taiwan
(111-2124-M-001-005, 112-2124-M-001-014,  112-2112-M-003-010-MY3, and 113-2124-M-001-008);
the Ministry of Education (MoE) of Taiwan Yushan Young Scholar Program;
the Physics Division, National Center for Theoretical Sciences of Taiwan;
the National Aeronautics and
Space Administration (NASA, Fermi Guest Investigator
grant 
80NSSC23K1508, NASA Astrophysics Theory Program grant 80NSSC20K0527, NASA NuSTAR award 
80NSSC20K0645); NASA Hubble Fellowship Program Einstein Fellowship;
NASA Hubble Fellowship 
grants HST-HF2-51431.001-A, HST-HF2-51482.001-A, HST-HF2-51539.001-A, HST-HF2-51552.001A awarded 
by the Space Telescope Science Institute, which is operated by the Association of Universities for 
Research in Astronomy, Inc., for NASA, under contract NAS5-26555; 
the National Institute of Natural Sciences (NINS) of Japan; the National
Key Research and Development Program of China
(grant 2016YFA0400704, 2017YFA0402703, 2016YFA0400702); the National Science and Technology Council (NSTC, grants NSTC 111-2112-M-001 -041, NSTC 111-2124-M-001-005, NSTC 112-2124-M-001-014); the US National
Science Foundation (NSF, grants AST-0096454,
AST-0352953, AST-0521233, AST-0705062, AST-0905844, AST-0922984, AST-1126433, OIA-1126433, AST-1140030,
DGE-1144085, AST-1207704, AST-1207730, AST-1207752, MRI-1228509, OPP-1248097, AST-1310896, AST-1440254, 
AST-1555365, AST-1614868, AST-1615796, AST-1715061, AST-1716327,  AST-1726637, 
OISE-1743747, AST-1743747, AST-1816420, AST-1935980, AST-1952099, AST-2034306,  AST-2205908, AST-2307887, AST-2407810); 
NSF Astronomy and Astrophysics Postdoctoral Fellowship (AST-1903847); 
the Natural Science Foundation of China (grants 11650110427, 10625314, 11721303, 11725312, 11873028, 11933007, 11991052, 11991053, 12192220, 12192223, 12273022, 12325302, 12303021); 
the Natural Sciences and Engineering Research Council of
Canada (NSERC); 
the National Research Foundation of Korea (the Global PhD Fellowship Grant: grants NRF-2015H1A2A1033752; the Korea Research Fellowship Program: NRF-2015H1D3A1066561; Brain Pool Program: RS-2024-00407499;  Basic Research Support Grant 2019R1F1A1059721, 2021R1A6A3A01086420, 2022R1C1C1005255, RS-2022-NR071771, 2022R1F1A1075115); the POSCO Science Fellowship of the POSCO TJ Park Foundation; NOIRLab, which is managed by the Association of Universities for Research in Astronomy (AURA) under a cooperative agreement with the National Science Foundation; 
Onsala Space Observatory (OSO) national infrastructure, for the provisioning
of its facilities/observational support (OSO receives funding through the Swedish Research Council under grant 2017-00648);  the Perimeter Institute for Theoretical Physics (research at Perimeter Institute is supported by the Government of Canada through the Department of Innovation, Science and Economic Development and by the Province of Ontario through the Ministry of Research, Innovation and Science); the Portuguese Foundation for Science and Technology (FCT) grants (Individual CEEC program – 5th edition, CIDMA
through the FCT Multi-Annual Financing Program for R\&D Units UID/04106, CERN/FIS-PAR/0024/2021, 2022.04560.PTDC); the Princeton Gravity Initiative; the Spanish Ministerio de Ciencia, Innovaci\'{o}n  y Universidades (grants PID2022-140888NB-C21, PID2022-140888NB-C22, PID2023-147883NB-C21, RYC2023-042988-I); the Severo Ochoa grant CEX2021-001131-S funded by MICIU/AEI/10.13039/501100011033; The European Union’s Horizon Europe research and innovation program under grant agreement No. 101093934 (RADIOBLOCKS); The European Union “NextGenerationEU”, the Recovery, Transformation and Resilience Plan, the CUII of the Andalusian Regional Government and the Spanish CSIC through grant AST22\_00001\_Subproject\_10; ``la Caixa'' Foundation (ID 100010434) through fellowship codes LCF/BQ/DI22/11940027 and LCF/BQ/DI22/11940030; 
the University of Pretoria for financial aid in the provision of the new 
Cluster Server nodes and SuperMicro (USA) for a SEEDING GRANT approved toward these 
nodes in 2020; the Shanghai Municipality orientation program of basic research for international scientists (grant no. 22JC1410600); 
the Shanghai Pilot Program for Basic Research, Chinese Academy of Science, 
Shanghai Branch (JCYJ-SHFY-2021-013); the Simons Foundation (grant 00001470); the Spanish Ministry for Science and Innovation grant CEX2021-001131-S funded by MCIN/AEI/10.13039/501100011033; the Spinoza Prize SPI 78-409; the South African Research Chairs Initiative, through the 
South African Radio Astronomy Observatory (SARAO, grant ID 77948),  which is a facility of the National 
Research Foundation (NRF), an agency of the Department of Science and Innovation (DSI) of South Africa; the Swedish Research Council (VR); the Taplin Fellowship; the Toray Science Foundation; the UK Science and Technology Facilities Council (grant no. ST/X508329/1); the US Department of Energy (USDOE) through the Los Alamos National
Laboratory (operated by Triad National Security,
LLC, for the National Nuclear Security Administration
of the USDOE, contract 89233218CNA000001); and the YCAA Prize Postdoctoral Fellowship. This work was also supported by the National Research Foundation of Korea (NRF) grant funded by the Korea government(MSIT) (RS-2024-00449206). We acknowledge support from the Coordenação de Aperfeiçoamento de Pessoal de Nível Superior (CAPES) of Brazil through PROEX grant number 88887.845378/2023-00. We acknowledge financial support from Millenium Nucleus NCN23\_002 (TITANs) and Comité Mixto ESO-Chile.

We thank
the staff at the participating observatories, correlation
centers, and institutions for their enthusiastic support.
This paper makes use of the following ALMA data:
ADS/JAO.ALMA\#2017.1.00841.V and ADS/JAO.ALMA\#2019.1.01797.V.
ALMA is a partnership
of the European Southern Observatory (ESO;
Europe, representing its member states), NSF, and
National Institutes of Natural Sciences of Japan, together
with National Research Council (Canada), Ministry
of Science and Technology (MOST; Taiwan),
Academia Sinica Institute of Astronomy and Astrophysics
(ASIAA; Taiwan), and Korea Astronomy and
Space Science Institute (KASI; Republic of Korea), in
cooperation with the Republic of Chile. The Joint
ALMA Observatory is operated by ESO, Associated
Universities, Inc. (AUI)/NRAO, and the National Astronomical
Observatory of Japan (NAOJ). The NRAO
is a facility of the NSF operated under cooperative agreement
by AUI.
This research used resources of the Oak Ridge Leadership Computing Facility at the Oak Ridge National
Laboratory, which is supported by the Office of Science of the U.S. Department of Energy under contract
No. DE-AC05-00OR22725; the ASTROVIVES FEDER infrastructure, with project code IDIFEDER-2021-086; the computing cluster of Shanghai VLBI correlator supported by the Special Fund 
for Astronomy from the Ministry of Finance in China;  
We also thank the Center for Computational Astrophysics, National Astronomical Observatory of Japan. This work was supported by FAPESP (Fundacao de Amparo a Pesquisa do Estado de Sao Paulo) under grant 2021/01183-8.

APEX is a collaboration between the
Max-Planck-Institut f{\"u}r Radioastronomie (Germany),
ESO, and the Onsala Space Observatory (Sweden). The
SMA is a joint project between the SAO and ASIAA
and is funded by the Smithsonian Institution and the
Academia Sinica. The JCMT is operated by the East
Asian Observatory on behalf of the NAOJ, ASIAA, and
KASI, as well as the Ministry of Finance of China, Chinese
Academy of Sciences, and the National Key Research and Development
Program (No. 2017YFA0402700) of China
and Natural Science Foundation of China grant 11873028.
Additional funding support for the JCMT is provided by the Science
and Technologies Facility Council (UK) and participating
universities in the UK and Canada. 
The LMT is a project operated by the Instituto Nacional
de Astr\'{o}fisica, \'{O}ptica, y Electr\'{o}nica (Mexico) and the
University of Massachusetts at Amherst (USA).
The IRAM 30 m telescope on Pico Veleta, Spain and the NOEMA interferometer on Plateau de Bure,
France are operated by IRAM and supported by CNRS (Centre National de la Recherche Scientifique, France), MPG (Max-Planck-Gesellschaft, Germany), and IGN (Instituto Geográfico Nacional, Spain).
The SMT is operated by the Arizona
Radio Observatory, a part of the Steward Observatory
of the University of Arizona, with financial support of
operations from the State of Arizona and financial support
for instrumentation development from the NSF.
Support for SPT participation in the EHT is provided by the National Science Foundation through award OPP-1852617 
to the University of Chicago. Partial support is also 
provided by the Kavli Institute of Cosmological Physics at the University of Chicago. The SPT hydrogen maser was 
provided on loan from the GLT, courtesy of ASIAA.

This work used the
Extreme Science and Engineering Discovery Environment
(XSEDE), supported by NSF grant ACI-1548562,
and CyVerse, supported by NSF grants DBI-0735191,
DBI-1265383, and DBI-1743442. XSEDE Stampede2 resource
at TACC was allocated through TG-AST170024
and TG-AST080026N. XSEDE JetStream resource at
PTI and TACC was allocated through AST170028.
This research is part of the Frontera computing project at the Texas Advanced 
Computing Center through the Frontera Large-Scale Community Partnerships allocation
AST20023. Frontera is made possible by National Science Foundation award OAC-1818253.
This research was done using services provided by the OSG Consortium~\citep{osg07,osg09}, which is supported by the National Science Foundation award Nos. 2030508 and 1836650.
Additional work used ABACUS2.0, which is part of the eScience center at Southern Denmark University, and the Kultrun Astronomy Hybrid Cluster (projects Conicyt Programa de Astronomia Fondo Quimal QUIMAL170001, Conicyt PIA ACT172033, Fondecyt Iniciacion 11170268, Quimal 220002). 
Simulations were also performed on the SuperMUC cluster at the LRZ in Garching, 
on the LOEWE cluster in CSC in Frankfurt, on the HazelHen cluster at the HLRS in Stuttgart, 
and on the Pi2.0 and Siyuan Mark-I at Shanghai Jiao Tong University.
The computer resources of the Finnish IT Center for Science (CSC) and the Finnish Computing 
Competence Infrastructure (FCCI) project are acknowledged. This
research was enabled in part by support provided
by Compute Ontario (http://computeontario.ca), Calcul
Quebec (http://www.calculquebec.ca), and the Digital Research Alliance of Canada (https://alliancecan.ca/en).

The EHTC has
received generous donations of FPGA chips from Xilinx
Inc., under the Xilinx University Program. The EHTC
has benefited from technology shared under open-source
license by the Collaboration for Astronomy Signal Processing
and Electronics Research (CASPER). The EHT
project is grateful to T4Science and Microsemi for their
assistance with hydrogen masers. This research has
made use of NASA's Astrophysics Data System. We
gratefully acknowledge the support provided by the extended
staff of the ALMA, from the inception of
the ALMA Phasing Project through the observational
campaigns of 2017 and 2018. We would like to thank
A. Deller and W. Brisken for EHT-specific support with
the use of DiFX. We thank Martin Shepherd for the addition of extra features in the Difmap software 
that were used for the CLEAN imaging results presented in this paper.
We acknowledge the significance that
Maunakea, where the SMA and JCMT EHT stations
are located, has for the indigenous Hawaiian people.

\end{acknowledgements}

\end{appendix}

\end{document}